\documentclass[twocolumn,superscriptaddress,prx]{revtex4-1}
\usepackage{graphicx,epstopdf,amsmath,amsfonts,amssymb,color,comment,cancel, bm}
\usepackage[vcentermath]{youngtab}				% Source http://www.ctex.org/documents/packages/math/youngtab.pdf
\usepackage[caption=false]{subfig}
\usepackage{soul}
\usepackage{hyperref}
\hypersetup{
    colorlinks=true,
    linkcolor=blue,
    citecolor=blue,
    filecolor=magenta,      
    urlcolor=blue,
}

\newcommand{\nc}{\newcommand}

\def\beq{\begin{equation}}
\def\eeq{\end{equation}}
\def\beqs{\begin{equation*}}
\def\eeqs{\end{equation*}}
\def\bsubeq{\begin{subequations}}
\def\esubeq{\end{subequations}}
\def\bpm{\begin{pmatrix}}
\def\epm{\end{pmatrix}}
\def\bit{\begin{itemize}}
\def\eit{\end{itemize}}
\def\ben{\begin{enumerate}}
\def\een{\end{enumerate}}
\def\btab{\begin{tabular}}
\def\etab{\end{tabular}}
\nc{\al}{\alpha}
\nc{\ga}{\gamma}
\nc{\de}{\delta}
\nc{\ep}{\epsilon}
\nc{\ze}{\zeta}
\nc{\et}{\eta}
\nc{\ka}{\kappa}
\nc{\la}{\lambda}
\nc{\rh}{\rho}
\nc{\si}{\sigma}
\nc{\ta}{\tau}
\nc{\up}{\upsilon}
\nc{\ph}{\phi}
\nc{\ch}{\chi}
\nc{\ps}{\psi}
\nc{\om}{\omega}
\nc{\Ga}{\Gamma}
\nc{\De}{\Delta}
\nc{\La}{\Lambda}
\nc{\Si}{\Sigma}
\nc{\Up}{\Upsilon}
\nc{\Ph}{\Phi}
\nc{\Ps}{\Psi}
\nc{\Om}{\Omega}

\nc{\sket}[1]{|#1\rangle}
\nc{\sbra}[1]{\langle#1|}
\nc{\ket}[1]{\left|#1\right\rangle}
\nc{\bra}[1]{\left\langle#1\right|}
\nc{\vev}[1]{\left\langle#1\right\rangle}
\nc{\svev}[1]{\langle#1\rangle}
\nc{\TO}[1]{\vev{\mathcal{T}#1}}
\nc{\scalarp}[2]{\left\langle#1|#2\right\rangle}
\nc{\sscalarp}[2]{\langle#1|#2\rangle}
\nc{\commut}[2]{\left[#1,#2\right]}
\nc{\anticommut}[2]{\left\{#1,#2\right\}}
\nc{\scommut}[2]{[#1,#2]}
\nc{\santicommut}[2]{\{#1,#2\}}
\nc{\Leq}{\leqslant}
\nc{\Geq}{\geqslant}
\nc{\unit}{1\!\!1}
\nc{\bk}{\mathbf k}
\nc{\MM}{\mathcal{M}}
\nc{\Int}{\int\frac{d^4l}{(2\pi)^4}}
\nc{\ev}{\text{eV}}
\nc{\gev}{\text{GeV}}
\nc{\tev}{\text{TeV}}
\nc{\ptl}{\partial}
\nc{\ov}{\overline}
\nc{\Neq}[1]{n^{eq}_{#1}}
\nc{\Yeq}[1]{Y^{eq}_{#1}}
\nc{\gammaeq}[1]{\gamma^{eq}_{#1}}
\nc{\CPV}{\mathcal P}
\nc{\bdim}[1]{\underline{#1}}

\begin{document}
 
\title{Quantum theory of an atom in proximity to a superconductor}
\author{Matthias Le Dall}
\author{Igor Diniz}
\altaffiliation[Current address: ]{Instituto de Ci\^{e}ncias Exatas, Universidade Federal Rural do Rio de Janeiro, 23890-000 Serop\'{e}dica, RJ, Brazil}
\affiliation{Department of Physics and Astronomy, University of Victoria, Victoria, British Columbia, Canada V8W 2Y2}
\author{Luis G.\ G.\ V.\ Dias da Silva}
\affiliation{Instituto de F\'{\i}sica, Universidade de S\~{a}o Paulo, Caixa Postal 66318, 05315--970 S\~{a}o Paulo, S\~{a}o Paulo, Brazil}
\author{Rog\'{e}rio de Sousa}
\affiliation{Department of Physics and Astronomy, University of Victoria, Victoria, British Columbia, Canada V8W 2Y2}
\date{\today}

\begin{abstract}
The impact of superconducting correlations on localized electronic states is important for a wide range of experiments in fundamental and applied superconductivity. 
This includes scanning tunneling microscopy of atomic impurities at the surface of superconductors, as well as superconducting-ion-chip spectroscopy of neutral ions and Rydberg 
states. Moreover, atomlike centers close to the surface are currently believed to be the main source of noise and decoherence in qubits based on superconducting devices. 
The proximity effect is known to dress atomic orbitals in Cooper-pair-like states known as Yu-Shiba-Rusinov (YSR) states, but the 
impact of superconductivity on the measured orbital splittings and optical/noise transitions is not known. 
Here we study the interplay between orbital degeneracy and particle number admixture in atomic states, beyond the usual classical spin approximation. 
We model the atom as a generalized Anderson model interacting with a conventional $s$-wave superconductor. 
In the limit of zero on-site Coulomb repulsion ($U=0$), we obtain YSR subgap energy levels that are identical to the ones obtained from the classical spin model.
When $\Delta$ is large and $U>0$, the YSR spectra is no longer quasiparticle-like, and the highly degenerate orbital subspaces are split according to their spin, orbital, and 
number-parity symmetry.  We show that $U>0$ activates additional poles in the atomic Green's function, suggesting an alternative explanation for the peak splittings recently 
observed in scanning tunneling microscopy of orbitally-degenerate impurities in superconductors. 
We describe optical excitation and absorption of photons by YSR states, showing that many additional optical channels open up in comparison to the nonsuperconducting case. 
Conversely, the additional dissipation channels imply increased electromagnetic noise due to impurities in superconducting devices. 
\end{abstract}

\maketitle

\section{Introduction}
\label{sec:intro}

Several recent advancements are renewing the interest in the interaction of atoms or impurities with superconductors. Recently, scanning
tunneling microscopy (STM) experiments detected for the first time the
orbital splitting of Yu-Shiba-Rusinov (YSR) subgap states induced by Mn \cite{Ruby2016} and Cr \cite{Choi2017} ions on the surface of superconducting Pb. 
A rich spatial structure was observed, reminiscent of the orbital structure of $s$, $p$ and $d$ atomic orbitals. The origin of the energy splitting of YSR states was proposed to 
be due to crystal-field splitting \cite{Fulde1970, Moca2008}. 

In the field of atomic physics the use of the superconducting ion chip \cite{Nirrengarten2006} 
is revolutionizing optical spectroscopy and the manipulation
of neutral ions \cite{Verdu2009, Bernon2013}, including Rydberg states
with large radial quantum number $n$ \cite{Hermann-Avigliano2014, Beck2016}. An interesting question is whether the proximity of the atom to the superconductor provides additional 
opportunities to achieve control of quantum information encoded in atomic states. 

Finally, there is mounting evidence that unidentified atomlike states close to the surface of superconductors are responsible for noise and decoherence of quantum
bits based on superconducting devices \cite{Lanting2014a, Kumar2016}. Several localized centers were claimed to be sources of
noise, including dangling bonds \cite{deSousa2007}, interface states \cite{Choi2009a}, adsorbed molecules \cite{Lee2014}, and molecular oxygen \cite{Wang2015, Kumar2016}.  All 
theories to date ignore the impact of the superconducting proximity effect on the localized states causing noise. However, such atomlike centers have never been detected using 
microscopic probes such as STM. Anderson's theorem \cite{Anderson1959} proves that \emph{spinless}  centers cannot give rise to subgap bound states in superconductors. Therefore, 
the characterization of subgap bound states in superconducting devices using STM may prove invaluable to identify the centers that give rise to flux (spin) noise. 

All these developments motivate the study of the superconducting
proximity effect on general atomic and molecular states. Usually, these
studies are done by taking the localized state to be a classical spin, the so-called Shiba model \cite{Shiba1968}. This shows that the magnetic moment
of the atom leads to the formation of Cooper-pair bound states with energy within the
superconducting gap, the so-called YSR states \cite{Rusinov1969,Yu1965,Shiba1968}. Under Shiba's  classical spin approximation the
impurity potential is taken to be a $\delta$ function in space, resulting in YSR states that are localized wavepackets of band orbitals of the superconductor \cite{Flatte1997a, 
Moca2008}. In
other words, the classical spin approximation ignores the native orbital
structure of the isolated atomic state. While this is certainly a good approximation for a short-ranged atomic impurity potential with a small number of bound states, it fails to 
give a satisfactory description of long-ranged states in many-electron impurities or Rydberg states in atoms. 

Here we present a quantum theory of
orbitally-degenerate YSR states, to show that they give rise to many more energy level splittings and transitions than previously anticipated on the basis of the simpler classical 
model.  The orbitally-degenerate localized state can be realized by an atom or an impurity, but to keep the terminology concise we refer to both as \emph{atoms}. The paper is 
organized as follows: Section \ref{sec:Model} describes our model, which
is based on the Anderson model for a spherically-symmetric hydrogenic
atom hybridized with a conventional ($s$-wave) superconductor. Section
\ref{sec:ZeroU} describes exact solutions of this model for the case of zero Coulomb
repulsion ($U=0$) and general orbital quantum number $l$. We show that the $U=0$ limit reproduces the energy level structure emerging from Shiba's classical spin model. Therefore, 
we conclude that the quantum model with $U>0$ provides an important generalization 
to non-quasi-particle-like YSR energies and eigenstates. 

Section \ref{sec:LargeDelta} describes the nature of the YSR states for arbitrary $U$
in the large-$\Delta$ regime, the so-called superconducting atomic limit \cite{Affleck2000, Bauer2007, Meng2009}.
These solutions become exact for arbitrary $U$ in the limit $\Delta\rightarrow \infty$. We present explicit calculations for three kinds of atoms: $s$-wave,
$p$-wave, and mixed $s$ and $p$. These cases demonstrate qualitative
differences from $U=0$ in that the Bogoliubov picture breaks down and
additional energy level splittings appear due to $U>0$ in the presence of electron number admixture. We use angular
momentum as the organizing principle and the Young tableaux formalism to
categorize the various eigenstates. We adapt the usual atomic spectroscopic notation ${^{2S+1}\!L_J}$
to keep track of the symmetries and of the electron number content of the YSR
states. 
%Addressing the $s,p$ atomic states in detail allow the extrapolation of some of our results to general orbital angular momentum $l$.

Section~\ref{sec:Green's function} describes the impact of orbital degeneracy on the atomic single particle Green's function and local density of states measured by STM. It 
demonstrates that $U>0$ gives rise to orbital splittings of the quasiparticle peaks, suggesting novel interpretations for the STM experiments. 

Section~\ref{sec:SelectionRules} describes the impact of orbital degeneracy and electron number fluctuation on the optical transitions between YSR states. 
It shows that superconductivity induces many more optical transitions between localized states, demonstrating that optical spectroscopy of atomic impurities can reveal much more 
about superconductors than previously anticipated on the basis of simpler models. 

Finally, Sec.~\ref{sec:DiscussionConclusion} describes our conclusions, and discusses their implications for fundamental and applied research with superconductors.

\section{The model}
\label{sec:Model}

We consider an atom hybridized with a conventional s-wave Bardeen-Cooper-Schrieffer (BCS) superconductor. 
We assume the atom's valence shell (radial quantum number $n$) is just below the Fermi level $E_F$, so that we can neglect all other shells. 
In order to give a quantum description for the behavior of the atom in a superconductor we use the Anderson+BCS model  \cite{Bardeen1957,Anderson1961,Fano1961},
\beq\label{eq: complete hamiltonian}
	{\cal H}={\cal H}_{\rm atom}+{\cal H}_{\rm BCS}+{\cal H}_{\rm hyb}.
\eeq
Here the valence-shell Hamiltonian is given by \cite{Lin1988}
\beq
	{\cal H}_{\rm atom}=\sum_{l} \xi_{l}N_{l}+\sum_{l,l'}\frac{U_{l,l'}}{2}N_{l}\left(N_{l'}-\delta_{l,l'}\right),
\label{himp}
\eeq
where $\xi_{l}=\epsilon_{l}-E_F$ and
$N_l=\sum_{m\sigma}d^{\dagger}_{lm\sigma}d_{lm\sigma}$ are respectively the single particle energies and the number operators for localized or ``bare'' atomic electrons with 
orbital angular momentum $l=0,1,\ldots,n-1$ ($E_F$ is the Fermi energy). The operators $d^{\dagger}_{lm\sigma}$ ($d_{lm\sigma}$) create (annihilate) the bare atomic states with 
azimuthal quantum number $m=-l,\ldots,l$, and spin $\sigma=\pm1/2$. The on-site Coulomb energy $U_{l,l'}$ models electron-electron repulsion.

In the spherical-wave basis, the conduction electron wave functions normalized in the volume $V=\frac{4}{3}\pi R^3$ are given by $\psi_{klm}(r)=\sqrt{2/R}k 
j_l(kr)Y_{lm}(\hat{r})$, where $j_l$(kr) is the spherical Bessel function, and $Y_{lm}(\hat{r})$ is the spherical harmonic. Transforming from the plane-wave basis to the spherical 
wave basis makes the BCS Hamiltonian take the form,
\begin{eqnarray}
	{\cal H}_{\rm BCS}&=&\sum_{k,l,m,\sigma}\left\{\xi_{k}c^\dagger_{klm\sigma}c_{klm\sigma}\right.\nonumber\\
	&&\left.-\Delta\sigma(-1)^m c^\dagger_{klm\sigma}c^\dagger_{kl-m-\sigma}+{\rm H.c.}\right\}.
\label{hbcs}
\end{eqnarray}
The operator $c^\dagger_{klm\sigma}\,(c_{klm\sigma})$ creates (annihilates) a conduction electron in 
the state $k,l,m,\sigma$ and energy 
$\xi_{k}=\epsilon_{k}-E_F$, with $\epsilon_{k}=\hbar^2k^2/2m^*$ where $m^*$ is the effective electron mass. The BCS model assumes that the energy gap $\Delta$ is nonzero only for 
electrons within a cutoff $\hbar\omega_c$ from the Fermi level, i.e. Eq.~(\ref{hbcs}) only includes electrons with  $|\xi_k|< \hbar\omega_c$.  

The final ingredient of our model is the hybridization Hamiltonian. For a spherically-symmetric atomic potential $V(r)$ we get,
\beq
	{\cal H}_{\rm hyb}=\sum_{k,l,m,\sigma} V_{kl}\,c^\dagger_{klm\sigma}d_{lm\sigma}+{\rm H.c.},
\eeq
with hybridization amplitude $V_{kl}=\bra{klm\sigma}V\ket{lm\sigma}$ independent of azimuthal quantum number $m$. As a result, the hybridization linewidth for each orbital state 
is given by 
\beq
\Gamma_{l}=\pi \rho_{l}(\epsilon_{l}) \left|V_{k_ll}\right|^2,
\eeq
where $k_{l}=\sqrt{2m^*\epsilon_{l}}/\hbar$ and 
$\rho_{l}(E)$ is the energy density for conduction-electron $l$ states \footnote{For a free electron gas, $\rho_{l}(E)=\frac{2(2l+1)R}{\pi} \sqrt{\frac{m^*}{2\hbar^2E}}$ for 
$l=0,1,2,\ldots,l_{{\rm max}}$, and $\rho_l(E)=0$ for $l>l_{{\rm max}}$. The maximum $l$ relates to $E$ according to $l_{{\rm max}}=\frac{2R}{\pi}\sqrt{\frac{2m^*E}{\hbar^2}}$.}. 
$\Gamma_{l}$ represents the inverse lifetime for an electron to decay from a localized atomic state due to hybridization with the conduction electrons. 

\section{Exact solution for $U=0$}
\label{sec:ZeroU}

The $U_{l,l'}=0$ case provides valuable insight into the problem of orbitally-degenerate atoms because it can be solved exactly. Moreover, $U_{l,l'}=0$ is a good approximation for 
certain impurities such as oxygen, which usually has its valence 2p shell nearly fully occupied in most host materials.

In the absence of Coulomb repulsion,  the total Hamiltonian (\ref{eq: complete hamiltonian}) can be written as a sum of decoupled Hamiltonians,
\beq
{\cal H}= \sum_{l=0}^{n-1} \sum_{m=-l}^{l}{\cal H}_{lm},
\eeq
with each decoupled ${\cal H}_{lm}$ mixing orbitals that are the time reversal of each other,
\begin{eqnarray}
{\cal H}_{lm}&=&\xi_{l} \left(d^{\dag}_{lm\uparrow}d_{lm\uparrow}+d^{\dag}_{l-m\downarrow}d_{l-m\downarrow}\right)\nonumber\\
&&+\sum_{k} \left[\xi_k \left(c^{\dag}_{klm\uparrow}c_{klm\uparrow}+c^{\dag}_{kl-m\downarrow}c_{kl-m\downarrow}\right)\right.\nonumber\\
&&-\Delta(-1)^m \left(c^{\dag}_{klm\uparrow}c^{\dag}_{kl-m\downarrow}+{\rm H.c.}\right)\nonumber\\
&&\left.+V_{kl}\left(c^{\dag}_{klm\uparrow}d_{lm\uparrow}+c^{\dag}_{kl-m\downarrow}d_{l-m\downarrow}+{\rm H.c.}\right)\right].
\label{hnlm}
\end{eqnarray}
Each ${\cal H}_{lm}$ can be interpreted as an independent $U=0$ s-wave atom, which
can be diagonalized exactly using the following Bogoliubov transformation \cite{Yoshioka2000, deSousa2009b}:
\begin{subequations}
\begin{eqnarray}
d_{lm\uparrow}&=&u^*_{l}\Upsilon_{lm\uparrow} +(-1)^m v_l \Upsilon^{\dag}_{l-m\downarrow}\nonumber\\
&&+\sum_k \left(u^*_{kl} \gamma_{klm\uparrow}+(-1)^mv_{kl} \gamma^{\dag}_{kl-m\downarrow}\right),\\
d^{\dag}_{l-m\downarrow}&=&-(-1)^mv^*_{l}\Upsilon_{lm\uparrow} + u_{l} \Upsilon^{\dag}_{l-m\downarrow}\nonumber\\
&& +\sum_k \left(-(-1)^mv^*_{kl} \gamma_{klm\uparrow}+u_{kl} \gamma^{\dag}_{kl-m\downarrow}\right),
\end{eqnarray}
\end{subequations}
where we use uppercase Upsilon ($\Upsilon^{\dag}_{lm\sigma}$) to denote a creation operator for a \emph{bound} subgap 
quasiparticle, and lowercase gamma ($\gamma^{\dag}_{klm\sigma}$) for a \emph{continuum} overgap quasiparticle.
%where $\Upsilon^{\dag}_{lm\sigma}$ (uppercase upsilon) and $\gamma^{\dag}_{klm\sigma}$ (lowercase gamma) are creation operators for \emph{bound} (subgap) and \emph{continuum} 
(over-gap) quasiparticles, respectively. 
The energy of the bound quasiparticle is  denoted by $E_{l}$, and is given by the positive root of 
\beq
E^{2}_{l}\left[1+\frac{2\Gamma_{l}}{\sqrt{|\Delta|^2-E^{2}_{l}}}\right]=\xi^{2}_{l}+\Gamma^{2}_{l},
\label{enleqn}
\eeq
which can be shown to be always subgap, i.e., $E_{l}<|\Delta|$. The continuum quasiparticles have the usual BCS energy $E_k=\sqrt{|\Delta|^2+\xi^{2}_{k}}$,  
and the Green's functions for conduction electrons acquire poles at both $\pm E_k$ and $\pm E_l$. 
After applying the Bogoliubov transformation, the Hamiltonian (\ref{hnlm}) becomes 
\beq
{\cal H}_{lm}={\cal H'}^{{\rm atom}}_{lm}+{\cal H'}^{{\rm cont}}_{lm},
\eeq
with 
\begin{eqnarray}
{\cal H'}^{{\rm atom}}_{lm}&=&E_{l}\left(\Upsilon^{\dag}_{lm\uparrow}\Upsilon_{lm\uparrow}+\Upsilon^{\dag}_{l-m\downarrow}\Upsilon_{l-m\downarrow}\right)\nonumber\\
&&+\left(\xi_{l}-E_{l}\right)\theta(-\xi_{l}), 
\label{hprimeimp}
\end{eqnarray}
the diagonalized Hamiltonian for the atomic bound states, and
\begin{eqnarray}
{\cal H'}^{{\rm cont}}_{lm}&=&\sum_{k}\Big[E_{k}\left(\gamma^{\dag}_{klm\uparrow}\gamma_{klm\uparrow}+\gamma^{\dag}_{kl-m\downarrow}\gamma_{kl-m\downarrow}\right)\nonumber\\
&&+\left(\xi_k-E_k\right)\theta(k_F-k)\Big],
\end{eqnarray}
the diagonalized Hamiltonian for the continuum states. 

From Eq.~(\ref{hprimeimp}) we obtain the values of the ``dressed atom" many-particle energy levels:
\beq
E_{{\rm atom}}(\left\{N_{\Upsilon l}\right\})=\sum_{l}\left[E_{l} N_{\Upsilon l}+\left(\xi_{l}-E_{l}\right)\theta(-\xi_l)\right],
\label{eimpnUpsilon}
\eeq
where $N_{\Upsilon l}=0,1,\ldots,2(2l+1)$ is the number of excited bound quasiparticles in the $l$-th multiplet. The degeneracy of each level is given by 
\beq
g_{l}(N_{\Upsilon l})=\bpm2(2l+1)\\N_{\Upsilon l}\epm.
\label{gln}
\eeq
Not all dressed atomic energy levels remain bound states; for a level to be a bound state, its total energy must lie beneath the continuum, which here is given by the 
total ground-state energy (atom plus conduction electrons) plus one continuum quasiparticle excitation with ${\rm Min}\{E_k\}=|\Delta|$. Therefore, the criteria for $E_{{\rm 
atom}}(\left\{N_{\Upsilon l}\right\})$ to be a bound state is given by
\beq
\sum_{l}E_{l}N_{\Upsilon l} <|\Delta|.
\label{criteriaboundstate}
\eeq
Since $E_{l}<|\Delta|$, it follows that each $l$ multiplet gives rise to at least one excited bound atomic level [which is itself $2(2l+1)$-fold degenerate]. But when $E_{l}\ll 
|\Delta|$ , the number of different bound excited energy levels may increase up to $2(2l+1)$ per $l$ multiplet. 

We now compare this result to the classical spin (Shiba) model \cite{Shiba1968}.  Its main assumption is a spin-dependent scattering Hamiltonian of the form
\beq
{\cal H}_{{\rm scatt}}=-\sum_{k,k', l,m,\sigma,\sigma'}J_l c^{\dag}_{klm\sigma}\bm{\sigma}_{\sigma\sigma'} c_{k' l m \sigma' }\cdot \bm{S}_{{\rm Shiba}},
\label{hscatt}
\eeq
with the atom's spin operator taken to be a fixed classical vector $\bm{S}_{{\rm Shiba}}$, pointing up or down along $z$. Under the latter assumption Shiba's model is also 
quadratic, so it can be diagonalized exactly \cite{Sakurai1970}. This results in $2(2l+1)$ bound quasiparticles per $l$, each with energy
\beq
E^{{\rm Shiba}}_{l}=|\Delta| \left|\frac{1-\left(J_l S_{{\rm Shiba}}\pi\rho_{lF}/2\right)^{2}}{1+\left(J_l S_{{\rm Shiba}}\pi\rho_{lF}/2\right)^{2}}\right|,
\label{elShiba}
\eeq
where $\rho_{lF}=\rho_l(E_F)$ is the $l$-electron density at the Fermi level.

\begin{figure}
\centering
\includegraphics[width = 0.48\textwidth]{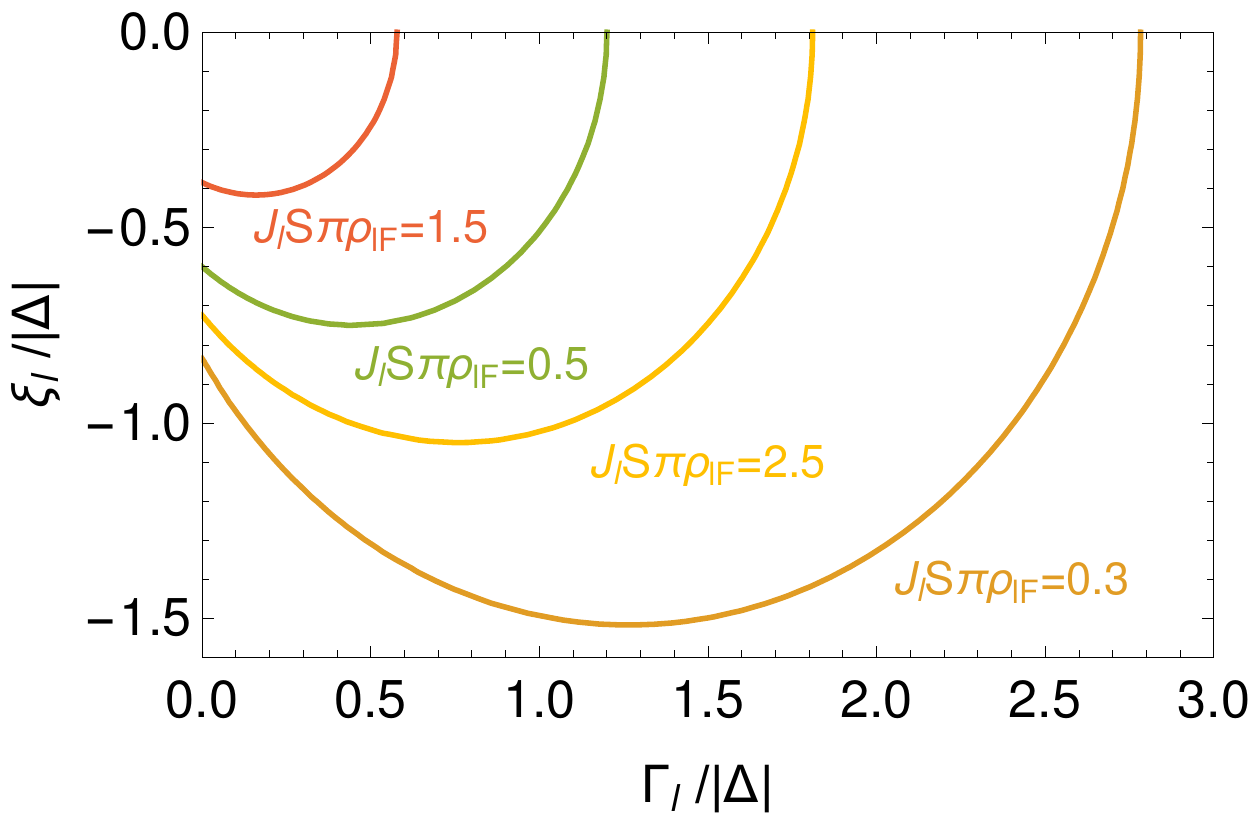}
\caption{(Color online) Relationship between the classical spin exchange scattering amplitude $J_l$ and the $U=0$ quantum model parameters $\xi_{l}, \Gamma_{l}$ 
divided by superconducting gap $\Delta$. For each $J_l$ the choice of $\xi_{l}, \Gamma_{l}$ in the corresponding curve makes the YSR energy levels of the classical Shiba model 
identical to the $U=0$ quantum YSR energy levels.}
\label{fig:correspquantumclass}
\end{figure}

It turns out that the bound quasiparticle energies of Eq.~(\ref{elShiba}) are equivalent to the ones from the quantum model in Eq.~(\ref{enleqn}), \emph{in the sense that for each 
given $J_lS_{{\rm Shiba}}$ one can find a set of $U=0$ Anderson model parameters $\xi_{l}, \Gamma_{l}$ that yield $E_{l}=E^{{\rm Shiba}}_{l}$}. This correspondence is shown in 
Fig.~\ref{fig:correspquantumclass}. We emphasize that 
the models are not equivalent: They differ qualitatively in that the Curie spin susceptibility is exactly equal to zero for the $U=0$ Anderson model, and is always nonzero for 
the Shiba model. The mapping that we propose is not obtainable from a canonical transformation such as that of Schrieffer and Wolff \cite{Salomaa1988}; it is instead a nonlinear 
map between parameters of two different models that makes the bound quasiparticle energies identical in both models. 

Therefore, we interpret the \emph{bound} energy levels emerging from the Anderson+BCS model [\ref{eq: complete hamiltonian}] as the quantum generalization of the 
YSR bound states obtained from the Shiba model. Figures~\ref{fig:YSRswavegamma},~\ref{fig: YSR state exact energies}~and~\ref{fig: optical transitions for s cross p ysr 
states} show the $U_{l,l'}=0$ YSR energy levels for $l=0$ (s), $l=1$ (p), and $l=0,1$ (s+p).  In the following sections we will study the impact of $U_{l,l'}>0$ in the YSR energy 
levels and eigenstates. 

\section{Approximate solution in the superconducting atomic limit (large $\Delta$)}
\label{sec:LargeDelta}

In the limit $|\Delta|\rightarrow\infty$ with $\hbar\omega_c/\Delta$ fixed, it is possible to integrate out the conduction electrons; the final result is an effective Hamiltonian 
that describes the proximity effect of superconductivity on undressed atomic operators \cite{Affleck2000, Bauer2007, Meng2009}. The exact diagonalization of the effective 
Hamiltonian yields the exact quantum YSR energy levels for $\Delta\rightarrow\infty$ and arbitrary $U, \xi_{l}, \Gamma_{l}$. In this limit all atomic energy levels satisfy 
inequality (\ref{criteriaboundstate}); therefore all of them are bound states. 

For $\Delta<\infty$, some of the higher energy atomic levels are not bound states, and the approach becomes an approximation, whose regime of validity is generally referred to as 
the \emph{superconducting atomic limit} (SCAL) \cite{Meng2009}. The reliability of the approximation can be checked {\it a posteriori}: If the SCAL result is a set of atomic 
energy levels $\{E_i\}$, the ones that can be considered ``reliable bound states'' are the ones satisfying $(E_i-E_0)\ll\Delta$. Therefore, the SCAL method is quite useful when one 
is interested in describing low energy excitations of atomic states in superconductors (i.e. up to $\sim 10$~GHz). 

In the first subsection below we give an explicit derivation of the effective Hamiltonian for the orbitally-degenerate case; the following subsections consider the particular 
cases 
of s, p, and mixed s+p atoms in detail. 

\subsection{Derivation of the large-$\Delta$ effective Hamiltonian}

The method relies on 
eliminating the conduction electron operators in the 
Heisenberg equation of motion for the atomic electron operators $d_{lm\sigma}(t)$. By writing the Heisenberg equation of motion for a conduction electron, $i\hbar\dot 
c_{klm\sigma}(t)=\commut{c_{klm\sigma}(t)}{{\cal H}}$, using the complete 
Hamiltonian (\ref{eq: complete hamiltonian}), and Fourier transforming to frequency $\omega$, we get
\beq
\begin{split}
	\left[(\hbar\omega)^2-E_{k}^2\right]c_{klm\sigma}(\omega)=&\Big[\left(\hbar\omega+\xi_{k}\right)V_{kl}\,d_{lm\sigma}(\omega)\\&+2\sigma\Delta 
V^*_{kl}\,d^\dagger_{lm\sigma}(-\omega)\Big].
\label{exactcommut}
\end{split}
\eeq

The key approximation of the SCAL method is to assume $\hbar\omega\ll|\Delta|$ in Eq.~(\ref{exactcommut}), so that its dependence on $\hbar\omega$ outside the argument of the 
operators is eliminated. This is equivalent to ``freezing out'' the dynamics of conduction electrons while allowing the dynamics of atomic electrons. As a result we get a 
linearized equation for the operators $c_{klm\sigma}(\omega)$ in terms of the atomic operators $d_{lm\sigma}(\omega)$. An inverse-Fourier transform yields
\beq
\begin{split}
	c_{klm\sigma}(t)=-\Big[&\frac{\xi_{k}}{E_{k}^2}V_{kl}d_{lm\sigma}(t)\\&+2\sigma(-1)^m\frac{\Delta}{E_{k}^2} V^{*}_{kl}d^\dagger_{l-m-\sigma}(t)\Big].
\label{cklmt}
\end{split}
\eeq
We plug Eq.~(\ref{cklmt}) into $\cal H_{\rm hyb}$ and $\cal H_{\rm BCS}$ to obtain the following effective Hamiltonian
\beq\label{eq: effective hamiltonian}
	{\cal H}_{\rm eff}={\cal H}'_{\rm atom}+{\cal H}_{\rm prox},
	\eeq
where ${\cal H}'_{\rm atom}$ is the atomic Hamiltonian with renormalized energy levels, and $\cal H_{\rm prox}$ accounts for 
the proximity effect on the atom,
\begin{subequations}
\begin{eqnarray}
	{\cal H}'_{\rm atom}&=&{\cal H}_{\rm atom}+\sum_{lm\sigma}\Sigma_{l}~d^\dagger_{lm\sigma}d_{lm\sigma},\\
	{\cal H}_{\rm prox}&=&-\sum_{lm\sigma}\sigma(-1)^{m}\tilde\Delta_{l}d^\dagger_{lm\sigma}d^\dagger_{l-m-\sigma} +{\rm H.c},\nonumber\\
\end{eqnarray}
\end{subequations}
where,
\bsubeq\label{eq: formal proximity potential}
\begin{align}
	&\tilde\Delta_{l}=\sum_{k}\frac{\Delta}{E_{k}^2}\left(V_{kl}^*\right)^{2},\label{eq: induced pairing}\\
	&\Sigma_{l}=-\sum_{k}\frac{\xi_{k}}{E^2_{k}}\left|V_{kl}\right|^{2},
\end{align}
\esubeq
are the induced pairing potential and the atomic self-energy, respectively. 

%\begin{figure}
%\includegraphics[width = 0.5 \textwidth]{proximity_potential_feynman}
%\caption{Feynman diagrams for processes giving rise to the impurity level self-energy corrections 
%(top), and SC pairing amplitudes (bottom). The dashed and double 
%lines represent the conduction electron normal and anomalous Green's functions, $G_{kl}(t)=-i\theta(t)\bra0c_{klm\sigma}(t)c^\dagger_{ 
%klm\sigma}(0)\ket0$, and $F_{kl}(t)=-i\theta(t)\bra0c_{klm\sigma}(t)c_{kl-m-\sigma}(0)\ket0$.}
%\label{fig: proximity potential feynman}
%\end{figure}

The renormalized atom Hamiltonian $\cal H'_{\rm atom}$ can be 
diagonalized by rotating the hydrogenic states into a set of operators $d'_{lm\sigma}$. From here on, we assume this has been done and keep the unprimed notation. Note 
that the induced pairing Eq.~\eqref{eq: induced pairing} does not bind atomic electrons that have different orbital angular momentum $l$. This is a consequence of our assumption 
of 
a spherically-symmetric atomic potential.

The proximity Hamiltonian ${\cal H}_{\rm prox}$ in Eq.~\eqref{eq: effective hamiltonian} describes an attractive force between atomic electrons resulting from the 
tunneling of Cooper pairs into the atom. This term induces pairing of the atomic state $m,\sigma$ with state 
$-m,-\sigma$.

The induced pairing $\tilde\Delta_{l}$ can be calculated explicitly in the BCS model, by converting $\sum_k$ into an integral over the energy range 
for nonzero $\Delta$:
\beq
\begin{split}
	\tilde\Delta_{l}=&\sum_{k}\frac{\Delta}{E_{k}^2}\left(V^*_{kl}\right)^{2}\\
	=&\int_{-\hbar\omega_c}^{\hbar\omega_c} d\epsilon_k\rho_l(\epsilon_k)\frac{\Delta}{E_{k}^2}\left(V^*_{kl}\right)^{2}\\
	=&~2\rho_{lF}\left(V^*_{kl}\right)^{2}\int_{0}^{\hbar\omega_c} d\xi_{k}\frac{\Delta}{\xi_{k}^2+\Delta^2}\\
	=&~2\rho_{lF}\left(V^*_{kl}\right)^{2}\tan^{-1}\left(\frac{\hbar\omega_c}{|\Delta|}\right)\frac{\Delta}{|\Delta|}.
\end{split}
\eeq
For most conventional superconductors we have $\hbar\omega_c\approx 30|\Delta|$ \cite{Tinkham1996}; in this regime, the induced pairing is well approximated by
\beq
	\tilde\Delta_{l}=\pi\rho_{lF}\left(V^*_{kl}\right)^{2}\frac{\Delta}{|\Delta|}=\Gamma_{l}\left(\frac{V^*_{kl}}{\left|V_{kl}\right|}\right)^{2}\frac{\Delta}{|\Delta|},
\label{tildednnl}
\eeq
which depends on the energy gap only through its phase. Therefore, within the SCAL approximation the magnitude of the proximity potential 
is always equal to the atom's hybridization linewidth, $|\tilde\Delta_{l}|=\Gamma_{l}$. 

\subsection{Effective Hamiltonian and labeling of atomic states}

To summarize, we may write the effective Hamiltonian (\ref{eq: effective hamiltonian}) as a sum of intra- and inter-$l$ contributions,
\beq
{\cal H}_{{\rm eff}}=\sum_{l}{\cal H}_{l}^{{\rm eff}} +\sum_{l<l'}{\cal V}_{ll'}.
\label{heff}
\eeq
The intra-$l$ Hamiltonian is given by
\begin{eqnarray}
{\cal H}_{l}^{{\rm eff}}&=& \xi_{l} N_{l}-\Gamma_{l}\sum_{m,\sigma}\left[\sigma(-1)^{m}d^{\dag}_{lm\sigma}d^{\dag}_{l-m-\sigma}+{\rm H.c.}\right]
\nonumber\\&&+\frac{1}{2}U_{l,l}N_l \left(N_l-1\right),
\end{eqnarray}
and the inter-$l$ one by 
\beq
{\cal V}_{ll'}=U_{l,l'} N_l N_{l'}.
\eeq
The effective Hamiltonian has the same symmetry as the total mother Hamiltonian (\ref{eq: complete hamiltonian}). In particular, with BCS s-wave pairing and a spherically 
symmetric atom, the model conserves total orbital momenta $\bm{L}=\sum_i\bm{L}_i$, and spin angular momenta, $\bm{S}=\sum_i\bm{S}_i$, as well as the total angular momentum 
$\bm{J}=\bm{L}+\bm{S}$. 
Therefore, it is convenient to represent the energy eigenstates using the set of commuting operators $\{\bm{S}^2, \bm{L}^2, \bm{J}^2, J_z\}$, leading to the usual spectroscopic 
notation 
${^{2S+1}\!L_{J}}$. 

Sometimes it is convenient to keep track of the number of electrons $N$ in a state; we do this by adding another superscript to the spectroscopy notation, ${^{2S+1}\!L^{N}_{J}}$. 
For example, the two-electron state with symmetry ${^{1}\!D_{2}}$ is written as $\ket{{^{1}\!D^{2}_{2}}}$. The proximity potential changes particle number by two, therefore we can 
separate the energy spectrum into states that contain either an even or an odd number of electrons [i.e., the Hamiltonian \eqref{heff} commutes with the number-parity operator]. 
The atom's energy eigenstates will contain mixtures of states with different numbers of electrons (either even or odd).

The dimension of the effective Hamiltonian basis is the number of ways to pick $N$ electrons out of all the single-electron orbitals, summed over all allowed $N$. For each $l$, 
the number of single-electron orbitals is $2(2l+1)$, so 
there are $g_{l}(N)$ basis states with $N$ electrons [see Eq.~(\ref{gln})]. 
The total dimension of the atom's $l$ multiplet is the sum over all allowed $N$,
\beq
	g_{l}^{{\rm total}}=\sum_{N=0}^{2(2l+1)}\bpm2(2l+1)\\N\epm=2^{4l+2},
\eeq
where we evaluated the sum using the binomial expansion for $(x+y)^{2(2l+1)}$ with $x=y=1$. Furthermore, applying the same expansion for $x=1$ and $y=-1$ we get that the number of 
states with even $N$ is equal to the number of states with odd $N$, 
\beq
g_{l}^{{\rm even}} = g_{l}^{{\rm odd}}=\frac{1}{2}g_{l}^{{\rm total}}= 2^{4l+1}.
\eeq

\subsection{YSR states from s-wave atom}

We start with the simple case of atomic $s$ states; while this case is quite elementary, it allows the introduction of our Young tableaux method \footnote{For an introduction to 
Young tableaux with one kind of orbital state see Sec.~6.5 of J.J. Sakurai, \emph{Modern Quantum Mechanics} (Addison-Wesley, Reading, MA, USA, 1994). Our method generalizes 
this technique to treat mixed spin and orbital states of different kinds.} to determine basis states that make our Hamiltonian block diagonal. 

The total number of $l=0$ orbitals is $2(2l+1)=2$; this means that atomic states can be made of zero, one, or two electrons, and that the total number of many-particle atomic 
states is $g_{s}^{{\rm total}}=2^{2}=4$.

\begin{figure*}[t!]
\begin{center}
\subfloat[]{\includegraphics[width=0.49\textwidth]{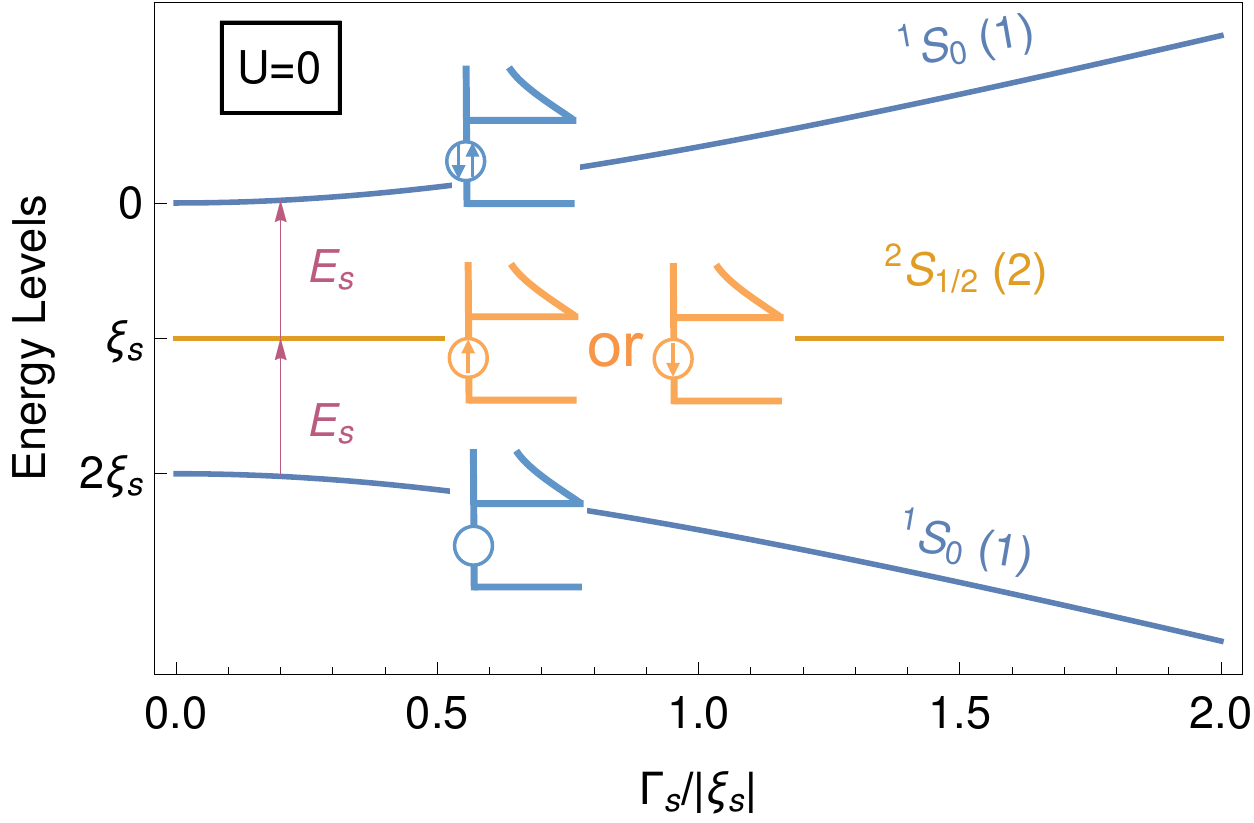}
\label{fig:YSRswavegamma}}
\subfloat[]{\includegraphics[width=0.51\textwidth]{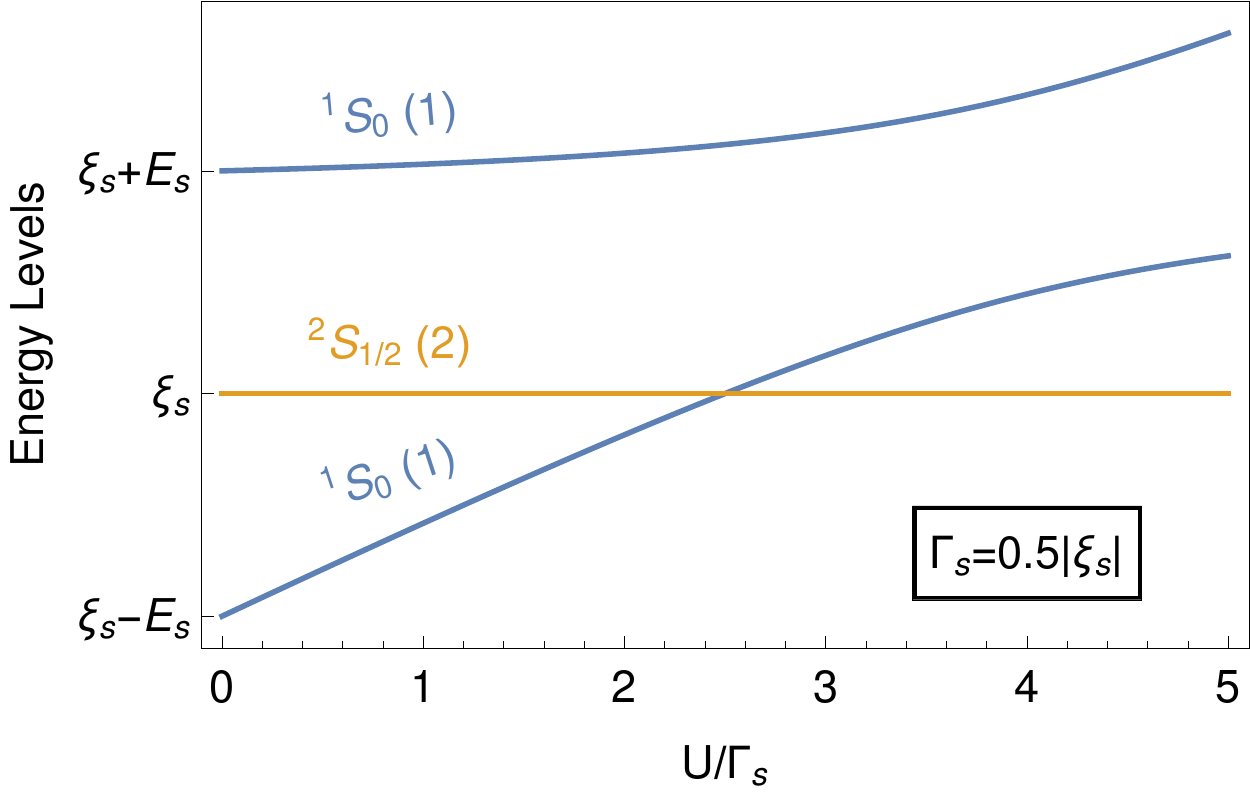}
\label{fig:YSRswaveU}}
\end{center}
\caption{(Color online) YSR spectra for $s$-wave atom. (a) Energy levels for $U=0$ as a function of proximity pairing $\Gamma_{s}/|\xi_{s}|$. Each many-particle level is 
represented by a quasiparticle local density of states. (b) Energy levels for $\xi_{s}=-2\Gamma_{s}$ and $U>0$, as a function of 
$U/\Gamma_{s}$. The multiplicity of each state ($2J+1$) is indicated in parentheses. Note how the quasiparticle picture breaks down (excited states do not differ by multiples of a 
single energy $E_s$), and the ground state changes from singlet to doublet at $U/\Gamma_{s}>2.5$.}
\label{fig:swaveenergies}
\end{figure*}

\subsubsection{Even-parity eigenstates}

The even-parity eigenstates can contain either 
zero (vacuum state $|0\rangle$) or two electrons, $g_{s}^{{\rm even}}=2$. 
We use an empty box to represent the spin state; the spin part of a two-electron basis state is determined from the tensor product of two boxes. In Young tableaux notation we have
\beq
\quad\yng(1)\otimes\yng(1)=\yng(2)\oplus\yng(1,1).
\eeq
The horizontal tableau is the symmetric (spin-triplet) state, while the vertical is the antisymmetric (spin-singlet) state. In the total spin representation (quantum number $S$) 
this equation would read 
$1/2\otimes1/2=1\oplus 0$.

Similarly, the orbital part of the basis state is found by using boxes filled with symbols. We use $\young(\bullet)$ to represent an $s$ orbital,
\beq
\quad\young(\bullet)\otimes\young(\bullet)=\young(\bullet\bullet)\oplus\cancel{\young(\bullet,\bullet)},
\eeq
which  in the total orbital representation (quantum number $L$) corresponds to $0 \otimes 0= 0$. 
Note that the vertical tableau was deleted because we cannot form an antisymmetric wave function out of one single particle orbital; in the Young tableaux method we cannot have a 
tableau
with the number of rows larger than the dimension of a box. 

The two-electron basis state is formed by the product of a symmetric orbital state times an antisymmetric spin state, leading to a state with symmetry ${^{1}\!S^{2}_{0}}$ (a 
singlet) in the 
${^{2S+1}\!L^{N}_{J}}$ notation:
\beq
	\young(\bullet\bullet)\otimes\yng(1,1)=\ket{{^{1}\!S^{2}_{0}}}=d^\dagger_{s\uparrow}d^\dagger_{s\downarrow}\ket0.
\eeq
This state, combined with the vacuum $\ket{0}$ (which is ${^{1}\!S^{0}_{0}}$) forms an optimal basis for even states. The proximity potential mixes both states leading to the 
effective Hamiltonian
\beq
	{\cal H}_{s}^{{\rm even}}=\bpm 0 &-\Gamma_{s}\\
	                                          -\Gamma_{s}&2\xi_{s}+U\epm
	\begin{array}{l}
		\ket{0}\\
		\ket{{^{1}\!S^{2}_{0}}}
	\end{array}.
\label{eq:hnssinglet}
\eeq

\begin{figure*}[t!]
\centering
\subfloat{\includegraphics[width=0.48\textwidth]{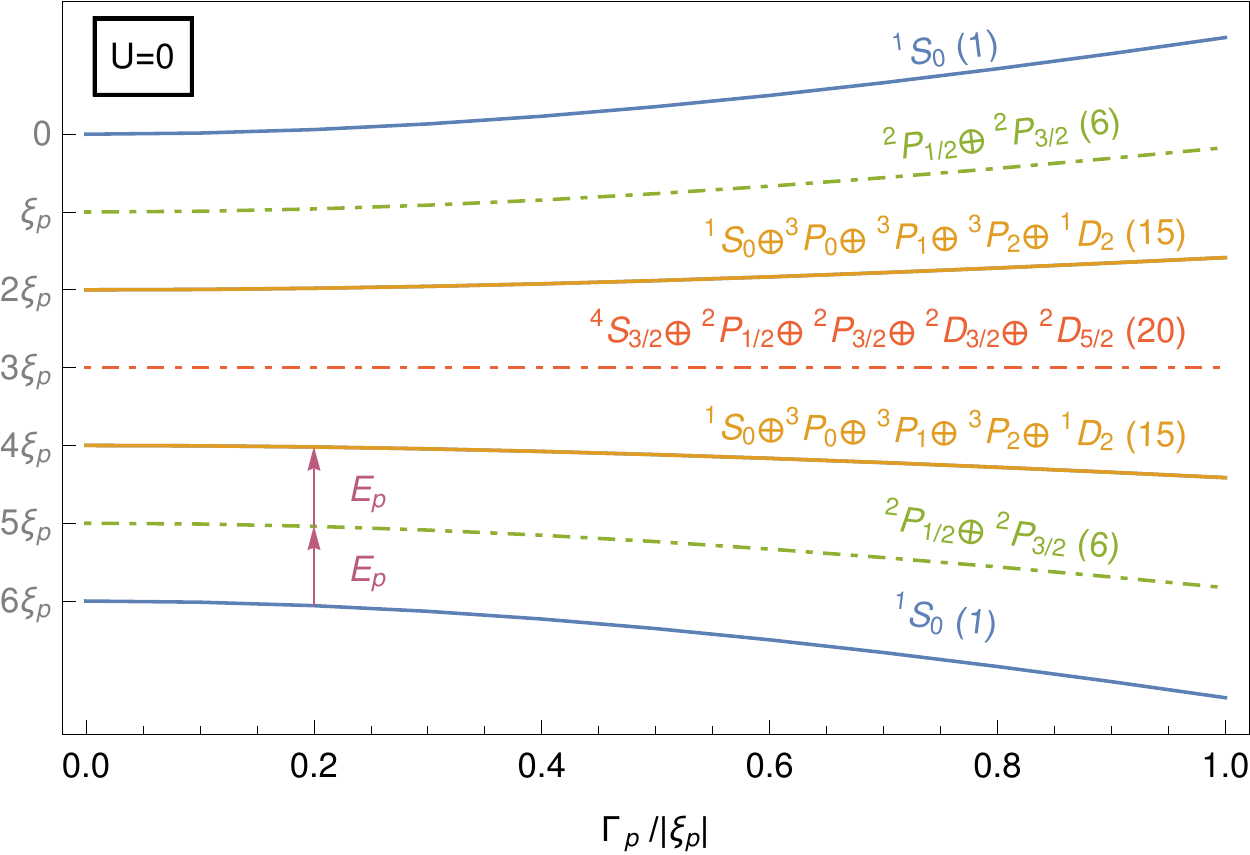}
\label{fig: YSR state exact energies}}
\subfloat{
\includegraphics[width=0.52\textwidth]{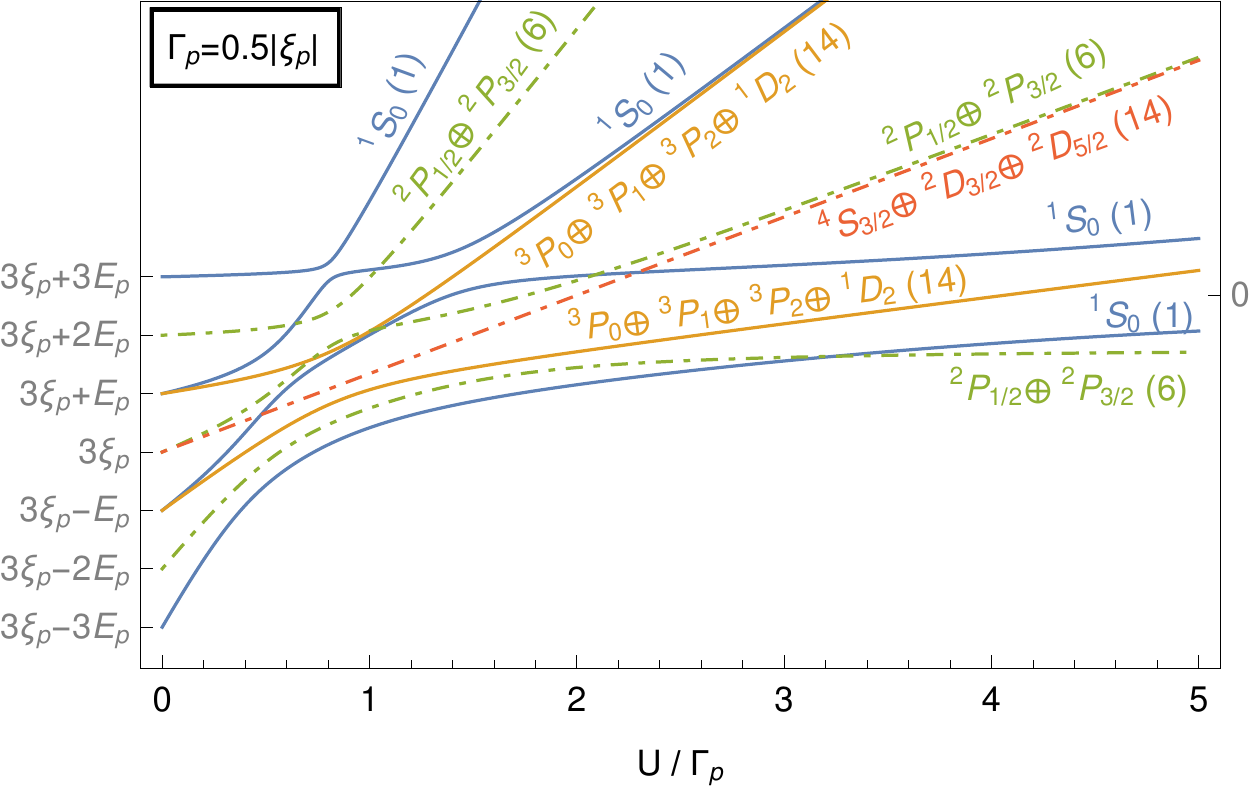}
\label{fig: YSR state exact energies coulomb}}
\caption[Orbitally-degenerate YSR spectrum]{(Color online) Orbitally-degenerate YSR energy levels for a $p$-wave atom. (a)  $U=0$ energies as a function 
of induced pairing $\Gamma_{p}/|\xi_{p}|$. (b) $U>0$ energies with induced pairing $\xi_{p}=-2\Gamma_{p}$ 
as a function of $U/|\xi_{p}|$.}
\label{fig: YSR energies and Coulomb}
\end{figure*}

\subsubsection{Odd-parity eigenstates}

There are $g_{s}^{{\rm odd}}=2$ one-electron states, with spin and orbital wave functions $\yng(1)$ and $\young(\bullet)$, respectively. The overall wave function is 
\beq
	\young(\bullet)\otimes\yng(1)=\ket{{^{2}\!S^{1}_{1/2}}}=d^\dagger_{s\uparrow}\ket0~{\rm or}~d^\dagger_{s\downarrow}\ket0.
\eeq
The proximity potential yields zero when applied to these states, because it attempts to add or remove two electrons; therefore our effective Hamiltonian is simply
\beq
{\cal H}^{{\rm odd}}_{s}=\bpm\xi_{s}\epm \ket{{^{2}\!S^{1}_{1/2}}},
\eeq
with degeneracy $2J+1=2$.

Figure~\ref{fig:YSRswavegamma} shows the dressed atomic energy levels for $U=0$ as a function of the proximity pairing energy $\Gamma_s$. We see that the only states that get 
affected by the superconducting proximity effect are the ${^{1}\!S_{0}}$ singlet states. The effect of superconductivity is to split them into \emph{bonding} and 
\emph{antibonding} Cooper-pair states. They get dressed by the superconductor so we refer to them as \emph{paired states}. For the usual situation of $\xi_{s}<0$ the bonding state 
is mainly $N=2$, while the antibonding state is mainly $N=0$. In contrast, the ${^{2}\!S_{1/2}}$ doublet states are not affected by superconductivity, so we refer to them as 
\emph{unpaired states}. The figure also shows the corresponding quasiparticle local density of states for each many-particle state; there we see that the unpaired states are 
obtained from a quasiparticle or quasihole excitation on top of the bonding ground state. Thus, the unpaired state corresponds to either both subgap states occupied or both empty. 
 
Figure~\ref{fig:YSRswavegamma} also shows the corresponding quasiparticle local density of states for each many-particle state; there we see that the unpaired states are obtained 
from a single quasiparticle excitation on top of the ground state. There are two kinds of quasiparticles (spin up or down), leading to two kinds of unpaired states. 

At low $U$ the ground state is the ${^{1}\!S_{0}}$ singlet; at larger $U$ a quantum phase transition occurs and the ${^{2}\!S_{1/2}}$ doublet becomes the ground state.

\subsection{YSR states from p-wave atom}

The next situation is that of an atom where only $p$-electrons are available to pair. 
Unlike the $s$-electron case, it is possible to form non-trivial states with odd numbers of electrons. The number of single-electron $p$-orbitals is $2(2l+1)=6$, so the total 
number of YSR states is $g_{p}^{{\rm total}}=2^6=64$. 

\begin{figure*}[t!]
\centering
\subfloat[]{\includegraphics[width=0.49\textwidth]{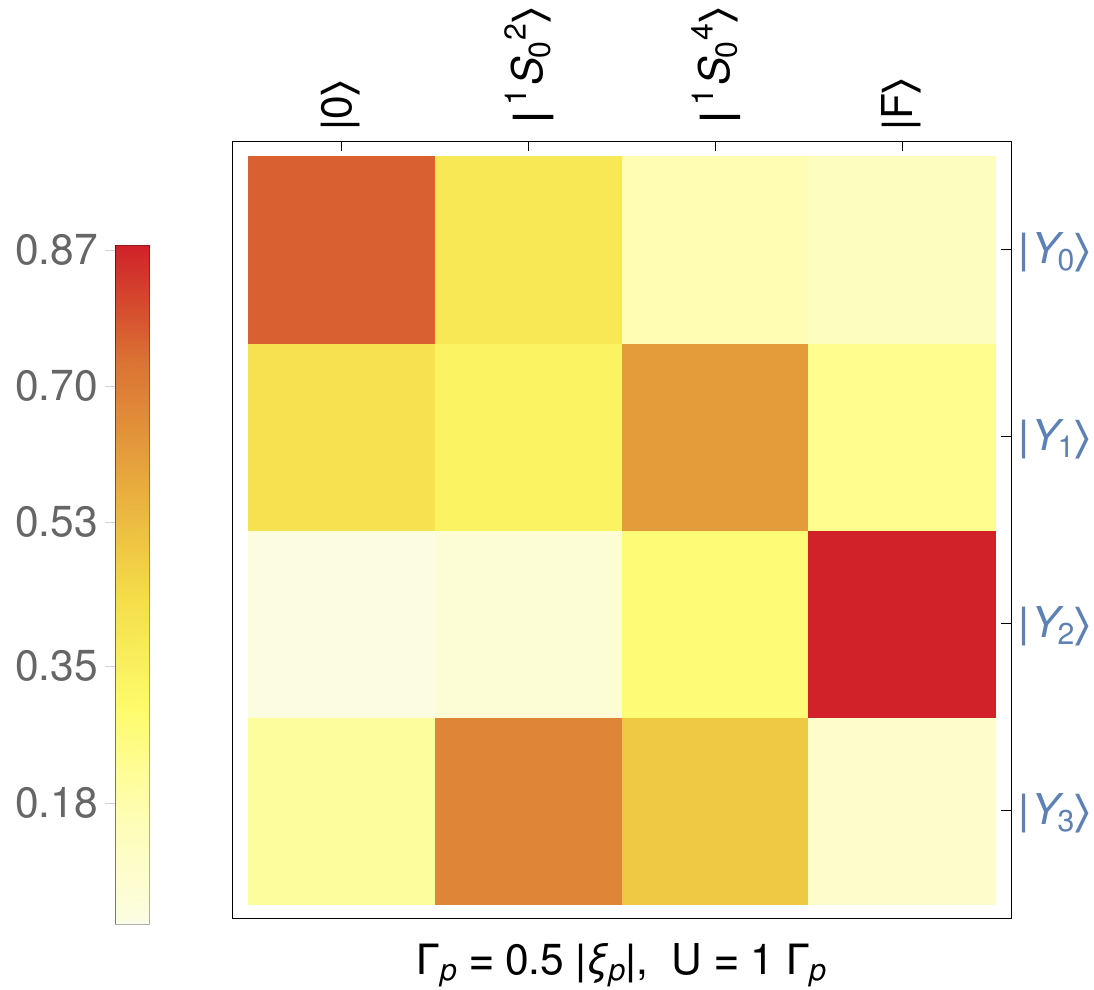}
\label{fig: singlet components low U}}
\subfloat[]{
\includegraphics[width=0.49\textwidth]{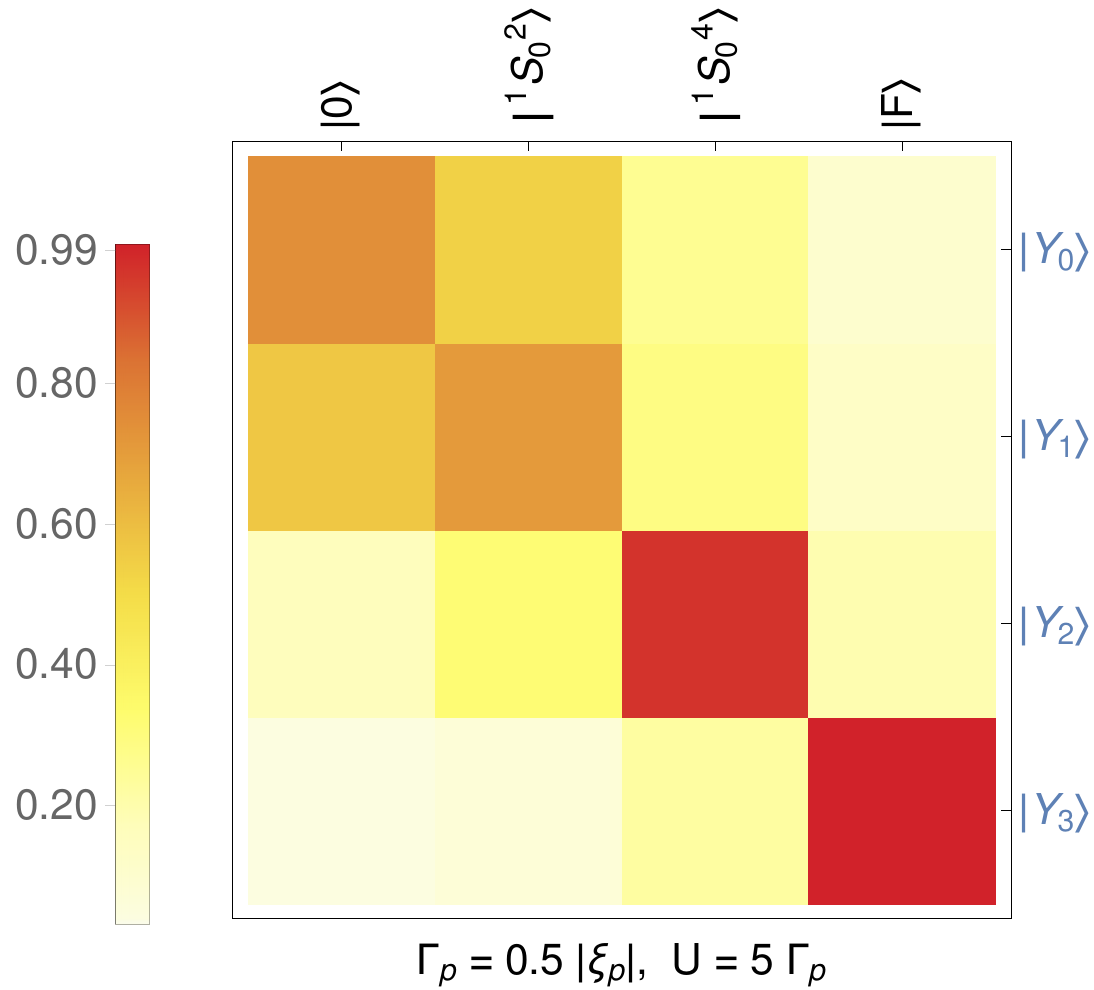}
\label{fig: singlet components large U}}
\caption[Components of YSR eigenstates]{
(Color online) Energy eigenstates of the $p$-wave atom, ${^{1}\!S_{0}}$ multiplet, at (a) low $U$ and (b) large $U$. The eigenstates are shown in the vertical axis, labeled 
$\sket{Y_{0,1,2,3}}$ in order of 
increasing energy. The horizontal axis shows basis states containing zero, two, four, and six electrons. The color diagram represents the probability of measuring each 
$N$-electron 
basis state.  At low $U$, all ${^{1}\!S_{0}}$ eigenstates display large particle-number fluctuation. At larger $U$ only the two lowest lying energy levels display particle-number 
fluctuation.}
\label{fig: singlet components}
\end{figure*}

\subsubsection{Even-parity eigenstates}

The number of even-numbered YSR states
is $g_{p}^{\rm even}=2^5=32$. The two-electron states are obtained 
by adding two electrons into the vacuum ${^{1}\!S^{0}_{0}}$; the four-electron states are obtained by removing two electrons from the maximally occupied state, 
$\ket{F}=\prod_{m,\sigma} 
d^{\dag}_{m\sigma}|0>$, which is also a singlet, ${^{1}\!S^{6}_{0}}$. In this way we find that the two-electron states have the same symmetries as the four-electron states.

The spin symmetries of the two-electron states are
\beq
	\yng(1)\otimes\yng(1)=\underbrace{\yng(2)}_{S=1}\oplus\underbrace{\yng(1,1)}_{S=0}, 
\label{youngtableax2espin}
\eeq
and using the tableau $\young(\circ)$ to represent a $p$ orbital we find their orbital symmetries to be
\beq
	\young(\circ)\otimes\young(\circ)=\underbrace{\young(\circ\circ)}_{L=0\oplus 2}\oplus\underbrace{\young(\circ,\circ)}_{L=1}.
\label{youngtableax2eorbit}
\eeq
Therefore, the two-electron basis states have the following symmetry:
\beq
\begin{split}
&\left(\young(\circ\circ)\otimes\yng(1,1)\right)\quad\oplus\quad\left(\young(\circ,\circ)\otimes\yng(2)\right)\\
&=\left({^{1}\!S_{0}}\oplus {^{1}\!D_{2}}\right)\oplus\left({^{3}\!P_{0}}\oplus{^{3}\!P_{1}}\oplus {^{3}\!P_{2}}\right).
\end{split}
\label{twoelectronpwave}
\eeq
The 4-electron states have exactly the same symmetries, and the even sector
Hamiltonian takes the block-diagonal form,
\beq
	{\cal H}^{\rm even}_{p}=
	\bpm
		{\cal H}_{\rm p}({^{1}\!S_{0}})&0&0\\
		0&{\cal H}_{\rm p}({^{1}\!D_{2}})&0\\
		0&0&\ddots\\
	\epm.
\eeq
It is a straightforward exercise to write all basis states explicitly. To do so, we  use a Clebsch-Gordan table to write the orbit and spin functions separately. 
Then we take their tensor product  and express the result in terms of $d^{\dag}_{m\sigma}$ operators applied on the vacuum. 
For example, consider the two-electron state with symmetry ${^{1}\!S_{0}}$. We know from Eq.~(\ref{youngtableax2espin}) that its $S=0$ came from $1/2\otimes 1/2$, and from 
Eq.~(\ref{youngtableax2eorbit}) that its $L=0$ came from $1\otimes 1$, so that we get
\begin{widetext}
\beq
\ket{{^{1}\!S^2_{0}}}=\ket{L=0,M_L=0}\ket{S=0,M_S=0}
=\frac{1}{\sqrt3}\left(d^\dagger_{1\uparrow}d^\dagger_{-1\downarrow}+d^\dagger_{-1\uparrow}d^\dagger_{1\downarrow}-d^\dagger_{0\uparrow}d^\dagger_{0\downarrow}\right)\ket0.
\eeq
After writing all the basis states we obtain explicit expressions for all blocks of the effective Hamiltonian.
For example, the singlet (${^{1}\!S_{0}}$) sector is given by
\beq
\begin{split}
	&{\cal H}_{p}({^{1}\!S_{0}})=\bpm
		0&\sqrt3\Gamma_{p}&0&0\\
		\sqrt3\Gamma_{p}&2\xi_{p}+U&-2\Gamma_{p}&0\\
		0&-2\Gamma_{p}&4\xi_{p}+6U&\sqrt3\Gamma_{p}\\
		0&0&\sqrt3\Gamma_{p}&6\xi_{p}+15U
	\epm
	\begin{array}{l}
	\ket0\\
	\ket{{^{1}\!S_{0}^{2}}}\\
	\ket{{^{1}\!S_{0}^{4}}}\\
	\ket{F}
\end{array}.
\end{split}
\label{hnp1s0}
\eeq
Note how the eigenstates of this matrix are mixtures of $N=0,2,4,6$ electrons. 
\end{widetext}

The other blocks (${^{1}\!D_{2}} \oplus {^{3}\!P_{0}} \oplus {^{3}\!P_1} \oplus {^{3}\!P_2}$) are all $2\times 2$ and can be expressed in a simple way:
\beq
\begin{split}
	&{\cal H}_{p}({^{2S+1}\!L_{J}})=\bpm
		2\xi_{p}+U&(-1)^J\Gamma_{p}\\
		(-1)^J\Gamma_{p}&4\xi_{p}+6U\\
	\epm
	\begin{array}{l}
	\ket{{^{2S+1}\!L_{J}^{2}}}\\
	\ket{{^{2S+1}\!L_{J}^{4}}}\\
\end{array}.
\end{split}
\eeq
Therefore, all states in the even sector are paired.

\subsubsection{Odd-parity eigenstates}

\begin{figure*}[t!]
\centering
\subfloat[]{\includegraphics[width=0.4\textwidth]{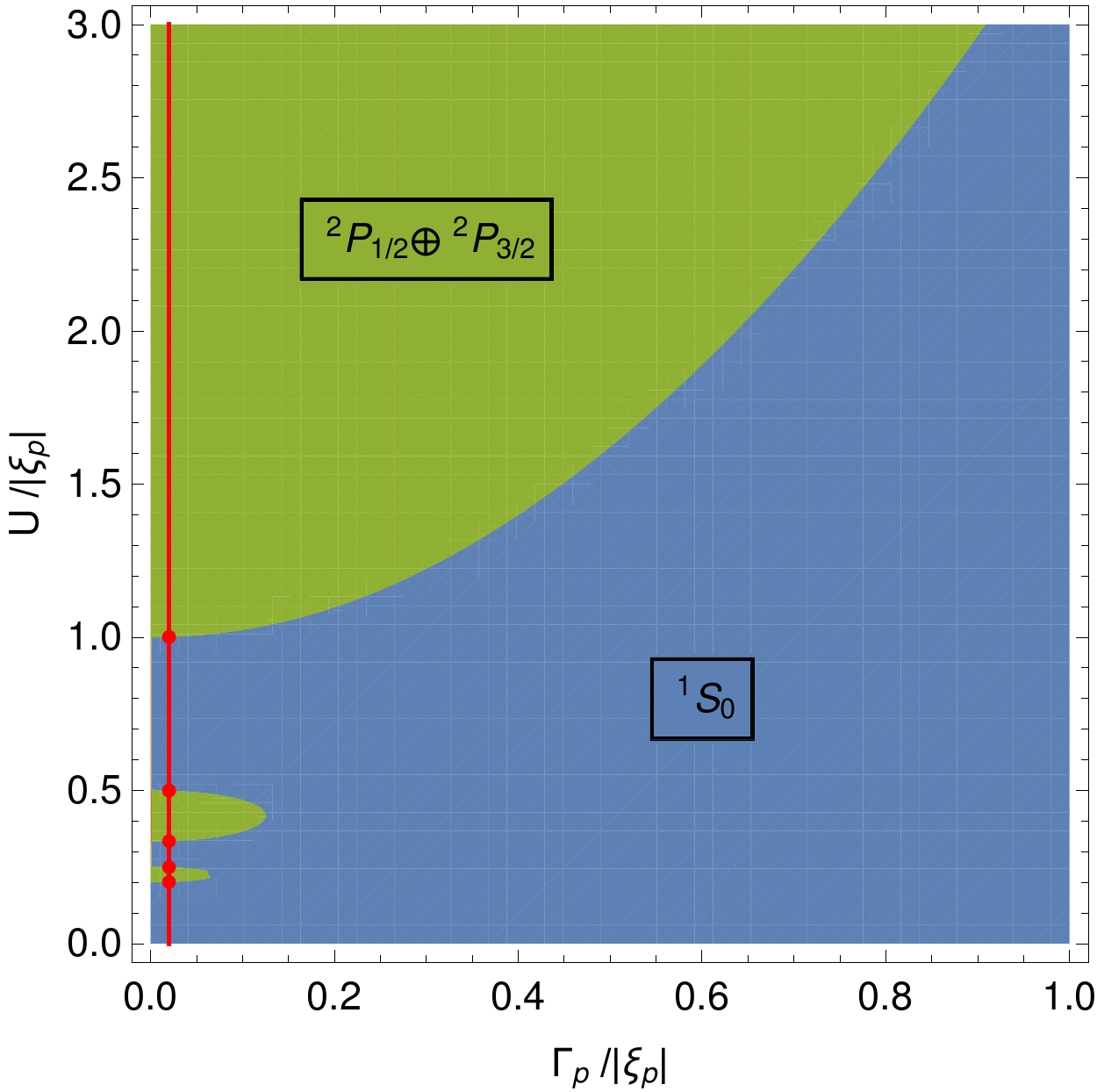}
\label{fig: ground state phase space}}
\subfloat[]{\includegraphics[width=0.61\textwidth]{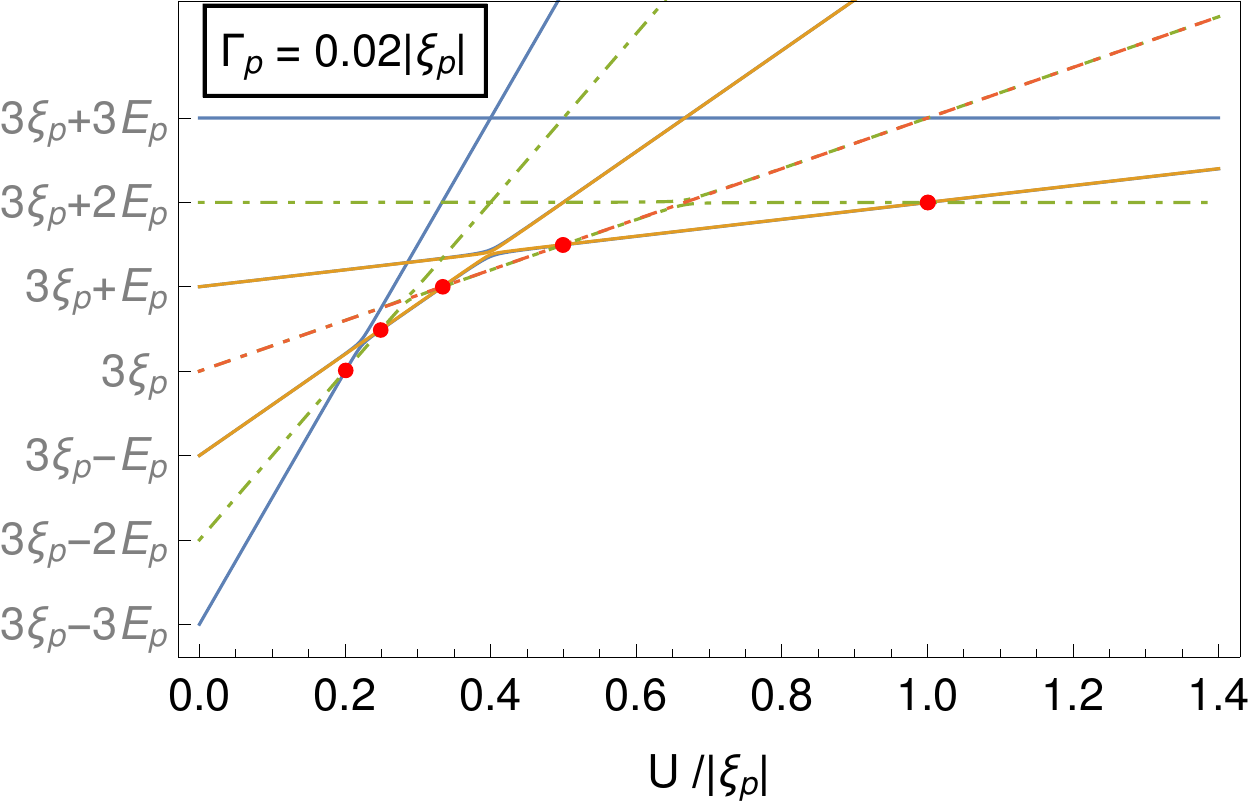}
\label{fig: even odd energy crossing}}
\caption[Parameter space of atomic ground state.]{(Color online) (a) Phase diagram for the ground state of the $p$-wave atom. 
The ground state alternates between the singlet ${^{1}\!S_{0}}$ (parity-even) and the sextuplet ${^{2}\!P_{1/2}}\oplus{^{2}\!P_{3/2}}$ (parity-odd). At $U\ll \xi_{p}$ we find a 
sequence of quantum phase transitions. 
(b) Energy levels for parameters along the red line in (a). The ground state transitions occur because of quantum fluctuations of particle number in each eigenstate.}
\label{fig: phase space and level crossing}
\end{figure*}

The number of odd-numbered states is also $g_{p}^{\rm odd}=32$.
The one-electron states are obtained by adding one electron to the vacuum state, and the 
five-electron states are obtained by removing one electron from $\sket{F}$. 
Their symmetries are ${^{2}\!P_{1/2}}\oplus~{^{2}\!P_{3/2}}$. 

The three-electron states are harder to obtain because they contain mixed spin-orbital symmetries.
Their spin and orbital parts are represented by the following Young tableaux:
\beq
\begin{split}
	&\yng(1)\otimes\yng(1)\otimes\yng(1)=\yng(3)\oplus\cancel{\yng(1,1,1)}\oplus\yng(2,1)\oplus\yng(2,1),\\
	&\young(\circ)\otimes\young(\circ)\otimes\young(\circ)=\young(\circ\circ\circ)\oplus\young(\circ,\circ,\circ)\oplus\young(\circ\circ,\circ)\oplus\young(\circ\circ,\circ).
\end{split}
\eeq
When combining orbit with spin we note that the orbital tableaux with mixed symmetry must combine with spin tableaux of the same kind of mixed symmetry, leading to
\beq
\begin{split}
&\left(\young(\circ,\circ,\circ)\otimes\yng(3)\right)\oplus\left(\young(\circ\circ,\circ) \otimes\yng(2,1)\right)\\
&=\left({^{4}\!S_{3/2}}\right)\oplus\left({^{2}\!P_{1/2}}\oplus~{^{2}\!P_{3/2}}\oplus~{^{2}\!D_{3/2}}\oplus~{^{2}\!D_{5/2}}\right),
\end{split}
\eeq
where we used
\beq
\young(\circ\circ,\circ)=1\oplus 2,\;\;\;\yng(2,1)=\frac{1}{2}.
\eeq

This results in identical matrices for the ${^{2}\!P_{J}}$ sectors,
\beq
\begin{split}
	&{\cal H}_{p}({^{2}\!P_{J}})=
	\bpm
		\xi_{p}&\sqrt2\Gamma_{p}&0 \\
		\sqrt2\Gamma_{p}&3\xi_{p}+3U&\sqrt2\Gamma_{p}\\
		0&\sqrt2\Gamma_{p}&5\xi_{p}+10U\\
	\epm
  \begin{array}{l}
	\sket{{^{2}\!P_{J}^{1}}}\\
	\sket{{^{2}\!P_{J}^{3}}}\\
	\sket{{^{2}\!P_{J}^{5}}}\\
  \end{array},
\end{split}
\eeq
leading to a total of $3(2+4)=18$ paired states.

In contrast, the other $14$ states ${^{4}\!S^{3}_{3/2}}\oplus {^{2}\!D^{3}_{3/2}}\oplus{^{2}\!D^{3}_{5/2}}$ remain unpaired with energy $3(\xi_{p}+U)$. 
It is easy to see that this occurs because these states have either maximum spin ($S=3/2$) or maximum orbital quantum number ($L=2$). Therefore, it is impossible to add two 
electrons to them without reducing either $S$ or $L$, so the action of ${\cal H}_{{\rm prox}}$ is always zero. Therefore, in contrast to the $s$-wave atom, some of the odd-parity 
eigenstates are paired. 

The complete energy spectrum of the $p$-wave atom is shown in Fig.~\ref{fig: YSR energies and Coulomb}. At $U=0$ 
we can represent each many-particle state as a diagram of occupied/unocupied subgap states (not shown), 
just as was done in Fig.~\ref{fig:YSRswavegamma}. The difference is that for the $p$-wave atom we have a total of six subgap states, three for spin up and three for spin down. As 
a result, three quasiparticle excitations are needed to reach the unpaired state with energy $3\xi_{p}$. For $U>0$ some representations are energy-split, and the quasiparticle 
picture breaks down. 
The energy splittings obey a weak version of Hund's rules, in the sense that it violates the hierarchy of the first and second rules. To see this, note that at each energy 
splitting,  states with \emph{either} larger $S$ or larger $L$ become lower in energy. The full hierarchy of Hund's rules can be incorporated by adding an exchange energy 
parameter to Eq.~(\ref{himp}) \cite{Lin1988}.

A graphical representation of the effect of the Coulomb interaction on the energy eigenstates is shown in Fig.~\ref{fig: singlet components}. 
There we see the amplitude of the 
mixing angles between the ${^{1}\!S_{0}}$ YSR energy eigenstates and basis vectors $\sket0,\sket{{^{1}\!S_{0}^{2}}},\sket{{^{1}\!S_{0}^{4}}},\sket{F}$. 
Figure \ref{fig: singlet components large U} shows that the singlet sector splits into blocks with increasing Coulomb interaction, 
with the largest-energy eigenstate mixing mostly with the maximally occupied state $\sket F$, and the next-to-largest eigenstate mixing mostly with the four-electron state. 
However, the two lowest-energy eigenstates remain mixtures of predominantly the vacuum and two-electron states. 

The Coulomb interaction leads to a level crossing between the lowest singlet ${^{1}\!S_{0}}$ and sextuplet ${^{2}\!P_{1/2}}\oplus~{^{2}\!P_{3/2}}$ energy levels. Thus, a quantum 
phase transition from singlet to sextuplet happens at $U>U_c$ where
\beq
	\frac{U_c}{|\xi_{p}|}\simeq1+3\frac{|\Gamma_{p}|^2}{|\xi_{p}|^2}.
\label{uc}
\eeq
The ground-state phase space as a function of $\Gamma_{p}/|\xi_{p}|$ and $U/|\xi_{p}|$ is shown in Fig.~\ref{fig: ground state phase space}. Note the multiple islands of sextuplet 
ground state. This is due to the fact that states with different symmetries have a different behavior as a function of the Coulomb interaction depending on the number of electrons 
they contain. This leads to level crossings between eigenstates of different symmetries, as is displayed in Fig.~\ref{fig: even odd energy crossing}, with parameters along the red 
line of the plot.

\subsection{YSR states from mixed s and p- wave atom}
\label{mixedsp}

It is interesting to consider the case of mixed s and p waves because it raises the possibility of optical excitation. For pure simplicity we will focus on the $U=0$ and large the 
$\Delta$ limit. As we have seen above the effect of $U$ is to decrease the energy of multiplets with large $S$ or $L$ quantum numbers. 

\begin{figure*}[t!]
\includegraphics[width=0.9\textwidth]{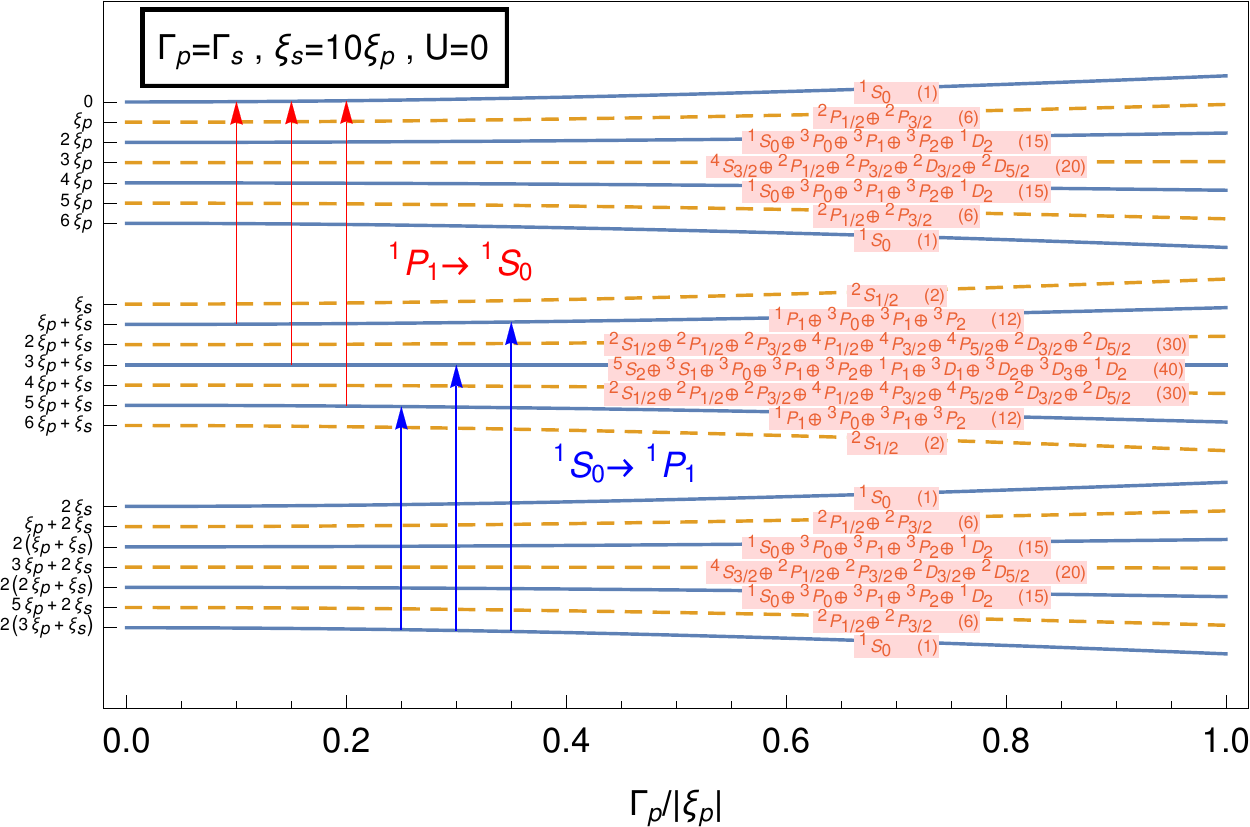}
\caption[Complete set of mixed s and p orbital eigenenergies]{(Color online) Complete set of mixed s and p energy levels at $U=0$. The solid (dashed) lines represent YSR states 
with an even (odd) 
number of electrons. The red arrows represent all one-photon transitions 
that have the vacuum as the final state, while the blue arrows are all one-photon transitions that have the ground state as the initial state. The pink labels represent the 
symmetry of each energy level, with the number in parentheses denoting the degeneracy out of a total of $256$ states. 
Note that all shown one-photon transitions are forbidden in the absence of the proximity effect. For example, at $\Gamma_p=0$ the ${^{1}S_{0}}\rightarrow{^{1}P_{1}}$ transitions 
are forbidden because they connect the eight-electron ground state to six, four, and two-electron states, respectively. When $\Gamma_p>0$, these states become admixtures of all 
even number of electrons, leading to the activation of the one-photon transitions.}
\label{fig: optical transitions for s cross p ysr states}
\end{figure*}

From Eq.~(\ref{hprimeimp})  we get the $sp$ Hamiltonian 
\beq
\begin{split}
	{\cal 
H}_{sp}=&E_{s}\sum_{\sigma}\Upsilon^\dagger_{s\sigma}\Upsilon_{s\sigma}+E_{p}\sum_{m,\sigma}\Upsilon^\dagger_{pm\sigma}\Upsilon_{pm\sigma}\\&+\left(\xi_{s}-E_{s}\right)+3\left(\xi_
{p}-E_{p}\right),
\end{split}
\eeq
with $E_{s}\approx\sqrt{\xi^2_{s}+\Gamma^2_{s}}$ and $E_{p}\approx\sqrt{\xi^2_{p}+\Gamma^2_{p}}$ in the large $\Delta$ limit. 
The spectrum is displayed in Fig.~\ref{fig: optical transitions for s cross p ysr states}.

Since the quasiparticle states have the same symmetry as their native orbital, the symmetry of the atomic states can be obtained from the same Young tableaux technique used above. 
With eight orbitals, the total number of atomic states is $g_{sp}^{{\rm total}}=2^8=256$. 

\begin{table}[t!]
\centering
\begin{tabular}{ccccc}
	\hline\hline
	Orbital& $\quad$ & Tableaux & Dimension & $L$\\\hline
	&&&&\\
	&&&\textit{Symmetric wave functions}&\\
	&&&&\\
	$l\otimes l'$ && $\young(\times\times)$ & $\underline{\bm{10}}=(4\times5)/2$ & $ S\oplus S\oplus P\oplus D$\\
	&&&&\\
	$s\otimes s$ && $\young(\bullet\bullet)$ & $\underline{\bm 1}=(1\times2)/2$ & $ S$\\
	&&&&\\
	$p\otimes p$ && $\young(\circ\circ)$ & $\underline{\bm 6}=(3\times4)/2$ & $ S\oplus D$\\
	&&&&\\
	$s\otimes p$ && $\young(\bullet\circ)$ & $\underline{\bm 3}=(1\times3)$ & $ P$\\
	&&&&\\
	&&&\textit{Antisymmetric wave functions}&\\
	&&&&\\
	$p\otimes p$ && $\young(\circ,\circ)$ & $\underline{\bm 3}=(3\times2)/2$ & $ P$\\
	&&&&\\
	$s\otimes p$ && $\young(\bullet,\circ)$ & $\underline{\bm 3}=(1\times3)$ & $ P$\\
	&&&&\\\hline\hline
\end{tabular}
\label{tab: antisymmetric mixed orbital wave function}
\caption{Set of all mixed orbital wave functions for two quasiparticles.}
\label{tab: symmetric and antisymmetric mixed orbital wave function}
\end{table}

Both the ground state $\sket{G}$ and the most excited state $\sket{F}$ are ${^{1}\!S_{0}}$ singlets, because they are the vacuum and completely filled quasiparticle states, 
respectively. The symmetry of the state with $N_s + N_p$ quasiparticles added to the vacuum is the same as the symmetry of the state with $N_s+N_p$ quasiholes added to $\sket{F}$. 

For brevity we will describe the symmetry of the two-quasiparticle states in detail; all the others can be obtained in an analogous fashion. There are a total of $28$ 
two-quasiparticle states (choose $2$ out of $8$). Their orbital wave function can be represented by mixed orbital notation $\young(\times)$, 
where $\times$ can assume one of four orbitals ($s$ or $m_p=-1,0,1$).  For two quasiparticles the orbital wave function becomes
\beq
\text{Young tableaux:}\quad\young(\times)\otimes\young(\times)=\young(\times\times)\oplus\young(\times,\times),
\eeq
which in dimensions reads $\underline{\mathbf4}\otimes\underline{\mathbf4}=\underline{\mathbf{10}}\oplus\underline{\mathbf6}$ \footnote{Here we used a simple rule for calculating 
the dimension of a Young tableau, $d=p/q$. To determine the numerator $p$:  Fill the top left box with the dimension of one index, 4 in this case. Fill in the rest of the tableau, 
increasing by one in each box to the right, and decreasing by one for each box down. Take the product of all the numbers to obtain $p$. To determine the denominator $q$: Fill each 
box with one plus the number of boxes to the right plus the number of boxes down. Take the product of all the numbers.}. 
A summary of all two-quasiparticle mixed orbital wave functions is found in Table~\ref{tab: symmetric and antisymmetric mixed orbital wave function}.
The total wave functions for the YSR states are the tensor products of the orbital and spin wave functions, in a way that is antisymmetric under electron interchange,
\begin{eqnarray}
	\young(\times\times)\otimes\yng(1,1)&=&{^{1}\!S_{0}}\oplus\left({^{1}\!S_{0}}\oplus {^{1}\!D_{2}}\right)\oplus {^{1}\!P_{1}},\\
	\young(\times,\times)\otimes\yng(2)&=&2\times \left({^{3}\!P_{0}}\oplus {^{3}\!P_{1}} \oplus {^{3}\!P_{2}} \right).
\end{eqnarray}
In an analogous fashion we can find the symmetry of the other states. All symmetries are shown in Fig.~\ref{fig: optical transitions for s cross p ysr states}.

\section{Atomic Green's function and local density of states}
\label{sec:Green's function}

The zero-temperature retarded Green's function
\beq
G^{R}_{m\sigma,m'\sigma'}(\epsilon)=\int_{0}^{\infty}dt \textrm{e}^{i\frac{\epsilon t}{\hbar}}\bra{G}\left\{d_{m\sigma}(t),d^{\dagger}_{m'\sigma'}(0)\right\} \ket{G},
\label{gmsmpsp}
\eeq
with $\ket{G}$ the ground state of the many-particle system, gives information about the energy required to add or remove one electron from the ground state. 
Equation~(\ref{gmsmpsp}) contains poles at $E_{\alpha}-E_{G}$, where $E_{\alpha}$ is an energy level of the many-particle system, ${\cal 
H}\ket{E^{i}_{\alpha}}=E_{\alpha}\ket{E^{i}_{\alpha}}$, with the superscript $i$ labeling the degenerate states. For $U=0$ these are just the quasiparticle energies. For $U>0$, 
the poles yield a generalized quasiparticle energy, in spite of the fact that for $U>0$ the atom's excited states can no longer be written as sums of quasiparticle energies [as 
was done in Eq.~(\ref{eimpnUpsilon})]. 

\subsection{Connection to STM experiments}

STM experiments measure a convolution of the local density of states defined by \cite{Yazdani1997a, Flatte1997a}
\beq
A(\epsilon)=-\frac{1}{\pi}{\rm lim}_{\eta\rightarrow 0^+}{\rm Im}\left\{\sum_{m,\sigma}G^{R}_{m\sigma,m\sigma}(\epsilon+i\eta)\right\}.
\label{defaepsilon}
\eeq
Plugging the energy eigenstates and eigenvalues into Eqs.~(\ref{gmsmpsp})~and~(\ref{defaepsilon})  leads to the spectral decomposition,
\begin{eqnarray}
A(\epsilon)&=&\sum_{\alpha,i,m,\sigma} \left\{
\left|\bra{E^{i}_{\alpha}}d^{\dagger}_{m\sigma}\ket{G}\right|^{2} \delta\left[\epsilon-(E_\alpha -E_{G})\right]\right.\nonumber\\
&&\left.+\left|\bra{E^{i}_{\alpha}}d_{m\sigma}\ket{G}\right|^{2} \delta\left[\epsilon+(E_\alpha -E_{G})\right]
\right\}.
\label{aepsilon}
\end{eqnarray}
We see that $A(\epsilon)$ with $\epsilon>0$ ($\epsilon<0$) represents the density for particle (hole) excitations, which is observed experimentally 
as a current from the atomic impurity to the STM tip at lower (higher) voltage. The height of the STM peak is given by the prefactor for each delta function:
\begin{subequations}
\begin{eqnarray}
w_{\alpha}^{+}&=& \sum_{i,m,\sigma}\left|\bra{E^{i}_{\alpha}}d^{\dagger}_{m\sigma}\ket{G}\right|^{2},\\
w_{\alpha}^{-}&=& \sum_{i,m,\sigma}\left|\bra{E^{i}_{\alpha}}d_{m\sigma}\ket{G}\right|^{2},
\end{eqnarray}
\end{subequations}
which are known as the particle ($+$) and hole ($-$) spectral weights, respectively.  Integrating Eq.~(\ref{aepsilon}) over all $\epsilon$ and using the closure relation leads to 
the sum rule,
\beq
\sum_{\alpha}\left(w_{\alpha}^{+}+w_{\alpha}^{-}\right)=2(2l+1).
\eeq

\subsection{Selection rules for nonzero spectral weights}

If the symmetry of the ground state is known, we are able to predict the set of energies $E_{\alpha}$ that can have nonzero spectral weights 
$w_{\alpha}^{\pm}$. There are two selection rules.

\subsubsection{Spin-orbit selection rule}

We want to find the allowed symmetries ${^{2S'+1}\!L'_{J'}}$ for the $\ket{E_{\alpha}}$ with nonzero matrix elements $\bra{E_{\alpha}}d^{\dagger}_{m\sigma}\ket{G}$, 
$\bra{E_{\alpha}}d_{m\sigma}\ket{G}$. The operators $d^{\dagger}_{m\sigma}$ and $d_{m\sigma}$ with single electron orbital angular momentum $l$ transforms like $L=l$, with 
symmetry ${^{1}\!l_{l}}\otimes {^{2}\!S_{1/2}} = {^{2}\!l_{|l+1/2|}}\oplus {^{2}\!l_{|l-1/2|}}$. If the ground state has symmetry ${^{2S+1}\!L_{J}}$, the Wigner-Eckart theorem 
constrains the possible symmetries to be
\beq
\left({^{2}\!l_{l+1/2}}\oplus {^{2}\!l_{|l-1/2|}}\right)\otimes {^{2S+1}\!L_{J}} = \sum_{S',L',J'}{^{2S'+1}\!L'_{J'}}.
\eeq
While the constraint on $J'$ is easy to find by simple addition of angular momenta,
\begin{eqnarray}
J'&=&\left||J-l|-\frac{1}{2}\right|, \left||J-l|-\frac{1}{2}\right|+1,\nonumber\\
&&\ldots,J+l+\frac{1}{2},
\end{eqnarray}
the analogous constraints on $S'$ and $L'$ are much more involved, requiring the application of the Young tableaux technique for each particular $l$ and ground state. 
 
\subsubsection{Parity selection rule}

The YSR eigenstate is a linear combination of either an even or an odd number of electrons. Since the operator $d^{\dagger}_{m\sigma}$ ($d_{m\sigma}$) adds (removes) one electron 
from $\ket{G}$, the local density of states will have peaks only for $\ket{E_{\alpha}}$ that have the \emph{opposite parity of $\ket{G}$}.

\begin{figure}[t!]
\includegraphics[width = 0.5 \textwidth]{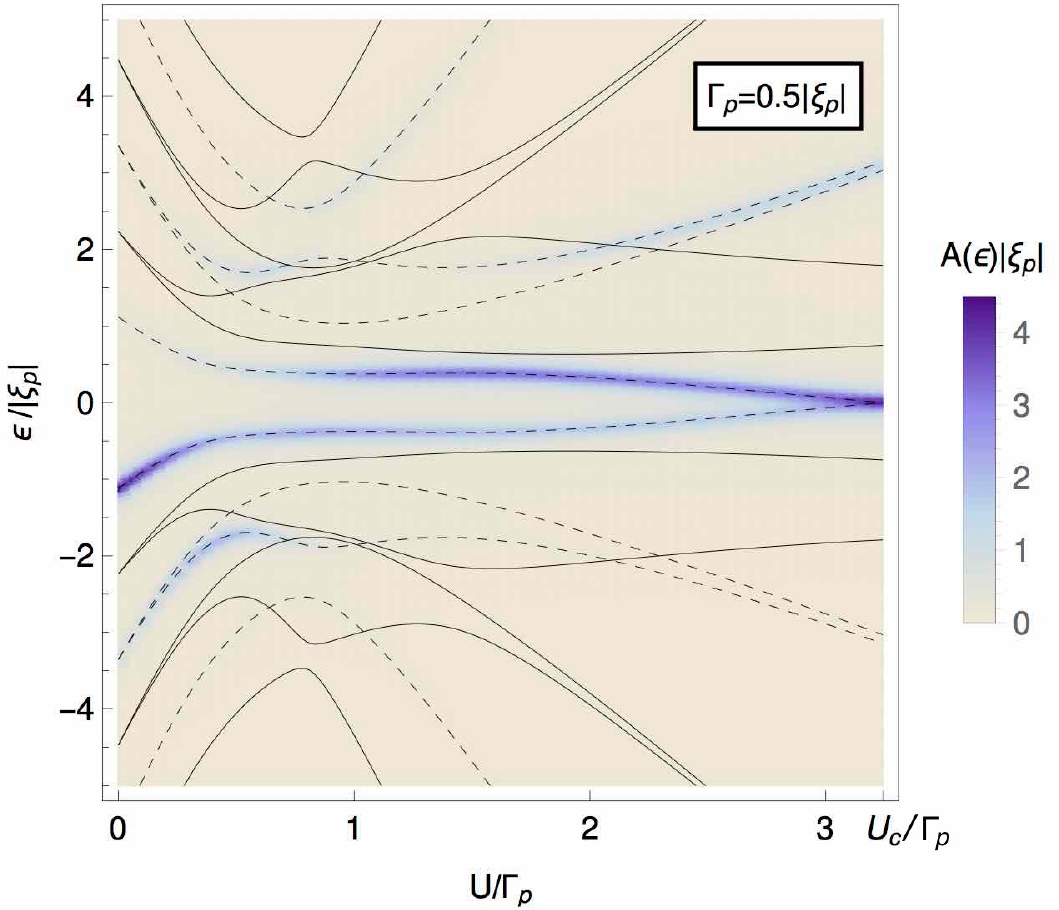}
\caption[]{(Color online) Local density of states for the $p$-wave atom, as a function of $U/\Gamma_p$ with $\xi_p<0$ and $\Gamma_p=0.5 |\xi_p|$. The $y$ axis represents the energy 
$\epsilon$ of the pole, with the spectral weight $A(\epsilon)$ given by the color map.
The expected location of each pole $E_\alpha-E_G$ is represented by black solid (dashed) lines for even-parity (odd-parity) states.}
\label{fig: p-wave local density of states}
\end{figure}

\subsection{Explicit calculation of local density of states for the $p$-wave atom}

The $p$-wave atom is in a ${^{1}S_{0}}$ singlet ground state for $U<U_c$ [see Eq.~(\ref{uc})]; therefore, its local density of states has peaks at the following excited states,
\beq
\left({^{2}\!P_{1/2}}\oplus {^{2}\!P_{3/2}}\right) \otimes {^{1}\!S_{0}} = {^{2}\!P_{1/2}}\oplus {^{2}\!P_{3/2}},
\eeq
which include all but one of the odd-parity energy levels. 
From Fig.~\ref{fig: YSR state exact energies coulomb} we expect a maximum of six nondegenerate poles, three positive and three negative. 
At $U>U_c$ the ground state changes to ${^{2}\!P_{1/2}}\oplus {^{2}\!P_{3/2}}$; as a result the 
poles that may get activated 
have the same symmetry as the two-electron states determined in Eq.~(\ref{twoelectronpwave}). 
These turn out to be the complete set of even-parity states. Thus, at $U>0$ up to twelve poles may get activated, six positive and six negative. 

Figure~\ref{fig: p-wave local density of states} shows the explicit numerical calculation of the local density of states as a function of $U/\Gamma_p$, for $0\leq U\leq U_c$. To 
plot the spectral density as a color map we represented the delta functions by Lorentzians with a small linewidth. The 
energy differences $E_\alpha- E_G$, which set the possible location of the poles, are represented by solid (dashed) lines for even-parity (odd-parity) $\ket{E_{
\alpha}}$.

At $U=0$ only two peaks are active, signaling the presence of a single bound quasiparticle. In contrast, \emph{at  $U>0$, we find that all the peaks allowed by symmetry get  
activated.} 

Our prediction of several activated peaks in the local density of states is in agreement with the recent observation of several subgap peaks in STM experiments of transition-metal 
impurities in superconducting Pb \cite{Ruby2016, Choi2017}. As we show here, the activation of multiple peaks and their splitting can occur solely due to the presence of Coulomb 
repulsion,  $U>0$, even in the absence of crystal-field splitting. 

\section{Optical selection rules for YSR states}
\label{sec:SelectionRules}

The previous sections showed that orbitally-degenerate YSR states may have nontrivial orbital, spin and total angular momentum 
symmetry. Electrons that can be in more than one orbital are able to absorb or emit one photon while transiting to a state with different orbital quantum number. 
Here we derive selection rules for optical excitation of YSR states. These selection rules are also relevant for electric noise, i.e., if two states are connected by the electric 
dipole operator, then the atomic impurity will emit noise at the frequency equal to the difference in energy between the two states. 

\subsubsection{Spin-orbit selection rule}

Under photon emission and absorption the atomic state with symmetry ${^{2S+1}\!L_{J}}$ may undergo an electric dipole transition to a state with 
different symmetry ${^{2S'+1}\!L'_{J'}}$, where \cite{Schwabl1995}
\beq
\begin{split}
	&L'-L =\pm1,\\
	&J'-J=0,\pm1\quad(J=0\rightarrow J'=0~{\rm forbidden}),\\
	&S'-S=0.
\end{split}
\eeq
All other transition rates (e.g. magnetic dipole, electric quadrupole) are smaller by a factor of $\alpha^2\sim 10^{-4}$.

\subsubsection{Parity selection rule}

Without loss of generality we focus on the case of Sec.~\ref{mixedsp} (mixed s and p orbitals). 
It is convenient to keep track of the content of the YSR states with the notation $s^ap^b$ where $a, b$ 
are the number of electrons in each single electron orbital. 

A YSR state can be in four parity configurations: even-even, odd-odd, even-odd and odd-even, referring to the parities of $a$ and $b$. 
Optical transitions connect  $s\leftrightarrow p$, which combined with the fact that $a+b$ must be conserved because dipole interactions conserve 
particle number, we arrive at the rule
\beq\label{eq: electron parity selection rule}
	s^ap^b~\xrightarrow[{\rm one~photon}]{}~s^{a\pm1}p^{b\mp1}.
\eeq
Unlike the dipole interaction, the induced proximity potential changes particle number by 2, and because the proximity potential does 
not mix s- with p-orbitals, the proximity effect mixes YSR states with different number of electrons but with the same parity. Thus a YSR state is the following mixture,
\beq
	\ket Y=\ket{s^ap^b}+\ket{s^{a\pm2}p^b}+\ket{s^ap^{b\pm2}}+\cdots
\eeq
Since every state in this series can undergo a transition according to Eq.~\eqref{eq: electron parity selection rule}, we deduce the parity selection rule,
\beq\label{eq: ysr parity selection rule}
\begin{split}
	{\rm even~a - even~b}~&\xrightarrow[{\rm one~photon}]{}~{\rm odd~a - odd~b},\\
	{\rm odd~a - even~b}~&\xrightarrow[{\rm one~photon}]{}~{\rm even~a - odd~b}.
\end{split}
\eeq
Note how this rule is much less restrictive than Eq.~(\ref{eq: electron parity selection rule}), indicating that a \emph{large number of optical excitation channels open up when 
the atom is in proximity to a superconductor}. 

The $sp$-mixed YSR eigenstates are shown in Fig.~\ref{fig: optical transitions for s cross p ysr states}, 
with solid (dashed) lines denoting even-parity (odd-parity) states. We can read off the electron content of each state by looking at the value of the energy level at 
$\Gamma_p=0$. For example, the ground state has energy
$2\xi_s+6\xi_p$ at $\Gamma_p=0$, showing that at $\Gamma_p>0$ it becomes a mixture of $0, 2$ $s$ electrons, and $0, 2, 4, 6$ $p$ electrons. 
With this information we can apply the above rules to identify allowed transitions. The total number of allowed transitions is very large, so here we 
focus on two particularly important sets of transitions: the ones that have the ground state as the initial state and the ones that have the vacuum state as the final state. 
Both the ground state and the vacuum have even-even parity and are pure ${^{1}\!S_{0}}$ singlets; thus we get 
\beq
\begin{split}
	{\rm even-even}~&\longleftrightarrow~{\rm odd-odd}\\
	{^{1}\!S_0}&\longleftrightarrow~{^{1}\!P_{1}}.
\end{split}
\eeq
As pictured in Fig.~\ref{fig: optical transitions for s cross p ysr states}, only three transitions satisfy those rules for both the vacuum and the ground state. 
It is interesting to point out that \emph{all shown transitions are forbidden in the nonsuperconducting case}.
Remarkably, the vacum state becomes optically active \emph{without the need for ionization}.  
 
\section{Discussion and conclusion}
\label{sec:DiscussionConclusion}

% Summary and quick description of the methods used
In summary, we showed that the proximity of an atomic state to a superconductor imprints Cooper-pairing behavior on the atomic states, with energy scale 
set by the rate of tunneling of atomic electrons into the superconductor. The resulting energy eigenstates are the orbitally-degenerate YSR states, which form generalized Cooper 
pairs containing mixtures of $0, 2, 4, \ldots$ (even-parity) or $1, 3, 5, \ldots$ (odd-parity) electrons. 

We resolved the nature of the YSR states using a combination of equation of motion, Bogoliubov canonical transformation, and effective Hamiltonian techniques. The eigenspectrum was 
calculated exactly in two extreme limits: (1) zero electron-electron repulsion $U$ and arbitrary energy gap $\Delta$, and (2)  $U>0$ and large $\Delta$. 

We described the YSR energy spectrum of atoms with mixed $s$- and $p$-wave character in detail, and organized their eigenstates according to their orbital, spin, and total angular momentum symmetries. The symmetries remain valid even in the regime when the Kondo effect wins the competition against Cooper pairing (when 
$\Delta<3k_BT_K$, with $T_K$ the Kondo temperature) \cite{Yoshioka2000, Bauer2007}.
Thus we argue that our results capture the essence of any physical situation with finite energy gap and Coulomb interaction, except for some renormalization of energy levels and 
matrix elements due to Kondo screening. The only qualitative limitation of our approach is that it does not allow the determination of which states are bound states in the small 
$\Delta$ regime. At $U=0$ our Eq.~(\ref{criteriaboundstate}) establishes a criterion for determining which atomic states remain bound when the atom is in proximity to a 
superconductor. 

We showed that nontrivial orbital symmetry combined with a particle-number admixture induced by the superconductor opens up several additional channels for optical emission and 
absorption. Therefore, optical spectroscopy of bound atoms or impurity states may reveal much more about the superconducting state than the usual spectroscopy of delocalized 
quasiparticles. The opening of additional decay and excitation channels is a general result that will also occur for other probes 
such as the ones based on magnetic transitions (e.g., spin resonance or Superconducting Quantum Interference Device (SQUID) microscopy) or electron tunneling (e.g., STM). 
Conversely, the fluctuation-dissipation theorem implies that impurities in superconductors will contribute additional electric, magnetic, and electron-number (charge) noise \cite{DiasdaSilva2015} at frequencies equal to the differences between their YSR energy levels.

While the number of allowed transitions is increased dramatically, some special states remain protected or long lived. These are the states that are not allowed by symmetry to  
form Cooper-pairs, i.e. they remain eigenstates of the electron-number operator [their energy level increases linearly with $U$; see, e.g., the fourteen-plet of Fig.~\ref{fig: YSR 
state exact energies coulomb}]. Out of these states one particular kind stands out, the ones with maximum spin $S_{{\rm max}}=(2l+1)/2$ and $L=0$ [see the ${^{4}\!S}_{3/2}$ in 
Fig.~\ref{fig: YSR state exact energies coulomb} and the ${^{5}\!S_{2}}$ in Fig.~\ref{fig: optical transitions for s cross p ysr states}]. 
All other atomic states have lower $S$, and the ground state has either $S=0$ (small-$U$ regime), or $S=1/2$ 
(large-$U$ regime where the average number of electrons in the ground state is close to one). Therefore, the $S_{{\rm max}}$ states are long lived, because their decay requires a 
cascade of transitions that are electric-dipole forbidden. 
 
%Relevance to recent STM experiments
Recently, STM experiments were able to detect for the first time the orbital splitting of atomic impurities in a superconductor \cite{Ruby2016,Choi2017}. 
The splittings, of the order of $1$ ~meV, were interpreted as arising from crystal-field splitting in the Shiba model \cite{Fulde1970, Moca2008}. 
Our Fig.~\ref{fig: p-wave local density of states} demonstrates that orbital splitting will occur solely due to electron-electron interaction, even in the absence of the crystal field.

Crystal-field splitting of bulk impurities in metals is expected to be negligibly small due to the screening of the lattice charge by the conduction electron gas; for atoms at  the surface the crystal-field splitting can reach values up to $0.3$~eV \cite{Herbst1977}. 
In contrast, transition-metal impurities are expected to have $U= 5-7$~eV (see Table 15-1 of \cite{Harrison2004}), with $U$ modified by as much as $1$~eV for impurities  near 
sample edges \cite{Miranda2016} or surfaces.
Therefore, several of the energy splittings detected by STM will be dominated by $U$.

% Relevance to Superconducting-based AMO
Recent experiments have used superconducting devices to serve as magnetic traps for Rydberg atoms above the surface, 
enabling precise control and manipulation of quantum information stored in the Rydberg atoms via laser excitations \cite{Hermann-Avigliano2014, Beck2016}.
While the proximity effect on these Rydberg states will decay exponentially as a function of their distance from the superconductor, the large Bohr radius of these states combined 
with improved trap design may allow the detection of the proximity effect on the atomic states. 
Our results show that extra optical channels open up as a result of the superconducting correlations, offering new avenues to manipulate 
quantum information stored in YSR-like states.

\begin{acknowledgments}
M. L. D., I. D.,  and R. d. S.  acknowledge financial support from NSERC (Canada) through its Discovery (RGPIN-2015-03938) and Collaborative Research and Development programs (CRDPJ 478366-14). 
L. G. D. S. acknowledges support from FAPESP Grant No. 2016/18495-4, CNPq Grants No. 307107/2013-2 and No. 449148/2014-9, and PRP-USP  NAP-QNano.
We acknowledge useful discussions with J. Fabian, K. J. Franke, and R.P.S.M. Lobo. 
\end{acknowledgments}

\bibliography{Orbital_YSR}
\end{document}